\documentclass{jfm}
\usepackage{graphics}
\usepackage{epstopdf, epsfig}
\usepackage{amsmath,mathrsfs,bm,color,float,mathtools}
\usepackage{url,verbatim}
\newcommand\beq{\begin{equation}}
\newcommand\eeq{\end{equation}}
\newcommand\bea{\begin{eqnarray}}
\newcommand\eea{\end{eqnarray}}
\renewcommand{\vec}[1]{\bm{#1}}

\newcommand{\blue}[1]{{\textcolor{black}{#1}}}

\shorttitle{Feedback control of invariant solutions}
\shortauthor{M. Linkmann, F. Knierim, S. Zammert, B. Eckhardt}

\title{Linear feedback control of invariant solutions in channel flow}

\author{
  Moritz Linkmann\aff{1}
  \corresp{\email{moritz.linkmann@ed.ac.uk}},
  Florian Knierim\aff{2},
  Stefan Zammert\aff{2},
  \\ \and 
  Bruno Eckhardt\aff{2}
  \aunote{Deceased on the $7^{\rm th}$ of August 2019.},
 }

\affiliation{
\aff{1}School of Mathematics and Maxwell Institute for Mathematical Sciences, 
University of Edinburgh, Edinburgh, EH9 3FD, United Kingdom \\
\aff{2}Fachbereich Physik, Philipps-University of Marburg,
D-35032 Marburg, Germany
}

\begin{document}

\maketitle

\begin{abstract}	
	Considering channel flow at Reynolds
	numbers below the linear stability threshold of the laminar profile as
	a generic example system showing a subcritical transition to turbulence
	connected with the existence of simple invariant solutions, we here
	discuss issues that arise in the application of linear feedback control
	of invariant solutions of the Navier-Stokes equations.  We focus on the
	simplest possible problem, that is, 
	travelling waves with one unstable direction.  In  view of
	potential experimental applicability we construct a pressure-based
	feedback strategy and study its effect on the stable, marginal and  unstable
	directions of these solutions in different periodic cells.  Even though the
	original instability can be removed, new instabilities emerge as
	the feedback procedure affects not only the unstable but also the
	stable directions.  We quantify these adverse effects and discuss their
	implications for the design of successful control strategies.  In order
	to highlight the challenges that arise in the application of feedback
	control methods in principle and concerning potential applications in
	the search for simple invariant solutions of the Navier-Stokes
	equations in particular, we consider an explicitly constructed analogue to
	closed-loop linear optimal control that leaves the stable directions
	unaffected.  
\end{abstract}


\section{Introduction}
\label{sec:intro}

Closed-loop control strategies such as linear optimal control \citep{Anderson90} 
are commonly used in engineering and industrial applications, fluid dynamics 
being only one example of such. 
In the present paper we consider linear feedback control as a means to stabilise 
exact nonlinear solutions of the Navier\textendash Stokes
equations, or, exact coherent structures, ECS.
ECS have been instrumental 
in the explanation of the subcritical transition to turbulence. In many shear flows 
the transition to turbulence occurs despite the linear stability of the laminar profile.
In pipe and plane Couette flow, for instance, the laminar profile is linearly stable at all Reynolds numbers. 
Plane Poiseuille flow becomes linearly unstable at a Reynolds number 
of $5772.22$ \citep{Orszag71b}, however, \blue{when subjected to finite-amplitude perturbations} the flow transitions much earlier. 
Exact coherent structures and their stability properties are not only of
interest to studies concerned with transitional flows. 
There is ample evidence
supporting the concept whereby the turbulent region of the state space of a
wall-bounded, parallel shear flow \blue{includes} many unstable ECS
\citep{Nagata1990,Hof2004,Hof2005,Duguet08,Duguet08b,Eckhardt07,Kawahara2012,Cvitanovic2013,Willis2016,Budanur17,Suri2017,Reetz2019,Reetz2020a,Reetz2020b},
with turbulence corresponding to a state-space trajectory travelling along the
ECS' stable and unstable manifolds resulting in frequent close passes to
different ECS. Once the state-space trajectory is in close vicinity of an ECS,
the properties of the turbulent state approximate those of that ECS.  Exact
solutions of the Navier\textendash Stokes equations can differ considerably in their
global and local properties, such as drag, mean profile or turbulence
intensity. The application of a feedback control procedure can be a useful
strategy to avoid states with undesirable properties such as high drag by
altering their stability properties, thus preventing state-space trajectories
to remain close to certain ECS or to confine the dynamics to certain
state-space volumes.  A dynamic feedback procedure based on adjustments of the
Richardson number succeeded in temporal stabilisation of otherwise transient
turbulent spots and stripes in stratified plane Couette flow
\citep{Taylor2016}.

A further potential application for feedback control in the context of ECS lies in the determination of 
so-called {\em edge states}, relative attractors on the {\em edge of chaos}, 
a \blue{codimension-1} 
manifold in state space that distinguishes between initial
conditions resulting in laminar or turbulent flow. The concept of edge states and edge manifolds 
is intrinsically connected to the transition to turbulence in 
many wall-bounded shear flows such as pipe, plane Couette and channel flow \citep{Itano01,Skufca06,Eckhardt07}.
Depending on the extent of the domain, edge states may be invariant solutions 
of the Navier\textendash Stokes equations or have chaotic 
dynamics and contain invariant solutions \citep{Budanur2018}. Edge states, 
or the invariant solutions contained therein, have by definition one 
unstable direction transversal to the edge \citep{Schneider07,Duguet08}, such that the dynamics will not
remain confined to it. The latter makes the determination 
of edge states, or invariant solutions therein, difficult. 
Bisection-based numerical methods
\citep{Itano01,Skufca06,Schneider07} are available, but
they are costly due to slow convergence
and high computational effort. Edge states can also be probed by minimal seed methods \citep{Pringle2010,Pringle2012,Pringle2015}, as
the smallest perturbation triggering turbulence, the {\em minimal seed}, is located infinitesimally close to the edge. 
It evolves along the edge, passes close to the edge state and eventually enters the turbulent region of state space. 

In small simulation domains or in \blue{symmetry-invariant} subspaces edge states are part
of an unstable lower branch of ECS 
that appear in a saddle-node bifurcation. 
In large domains, when edge states are chaotic and can contain 
ECS \citep{Budanur2018}, lower-branch ECS can be found within the edge state.
This suggests
that low-dimensional feedback stabilisation methods could be used to remove the
effect of the unstable directions, such that the edge state, 
or an invariant solution therein, is stabilised.  In
pipe flow, a simple feedback control strategy, where the Reynolds number is
adjusted in response to an observable connected with deviations from laminar
flow, indeed stabilises the dynamics to remain on the edge \citep{Willis17}.
Forward integration of the controlled system converged to previously unknown
edge states in form of travelling waves. For more complicated edge states such
as relative periodic orbits or those with chaotic dynamics, the controlled simulations
converged to objects in the vicinity of ECS of the uncontrolled system.

Here, we focus on feedback strategies in channel flow at Reynolds numbers below the linear 
stability threshold, as an example system showing a subcritical transition to turbulence. 
In order to highlight and discuss the challenges that arise in the application of linear feedback
control for the stabilisation of exact coherent structures, we attempt to stabilise some of 
simplest invariant solutions in minimal flow units, that is, edge states in form of travelling waves. 
Unlike \citet{Willis17} we aim to stabilise {\em known} invariant solutions.
To do so, we construct simple linear feedback procedures 
that are either (i) pressure-based and thus one step closer to experimental conditions, or 
(ii) adjoint-based and act on the single unstable direction by construction. 
In the first case, we monitor the controller's effect 
on global observables such as turbulent kinetic energy and skin friction coefficient, and we  
find that the controlled dynamics approaches values of these observables that correspond 
to the target states, however, the target states themselves are not stabilised. 
The reason lies in the occurrence of a new instability that is induced by the coupling of the control 
procedure to the edge states' stable directions. 
The second method removes such secondary instabilities by construction, 
however, care must be taken in its application in terms of the type of target state and 
the choice of global observable.  
Here, only a highly symmetric de-localised travelling wave has been successfully stabilised 
with this method, which illustrates the limitations of global one-dimensional feedback.

We begin with an introduction to the concept of
linear feedback control in sec.~\ref{sec:theory} in the context of invariant solutions, where the procedure is
explained and its effect is illustrated in low-dimensional examples. In
sec.~\ref{sec:control_nse} we use the general formalism outlined in
sec.~\ref{sec:theory} to develop the 
control strategies. 
Before applying the control procedures to 
the aforementioned edge states in direct numerical simulations of channel flow, 
we summarise the numerical details and describe the target states in sec.~\ref{sec:numerics}. 
Section \ref{sec:stabilisation} contains the main investigation into
stabilisation of edge states including the effect of the feedback control on
the stable directions. 
We summarise our results in sec.~\ref{sec:conclusions} alongside a discussion of the challenges that need to be overcome in 
the design of successful control strategies in the context of simple invariant solutions of the Navier\textendash Stokes equations.

\section{Stabilisation and control}
\label{sec:theory}
\begin{figure}
	\centerline{\includegraphics[width=.5\columnwidth]{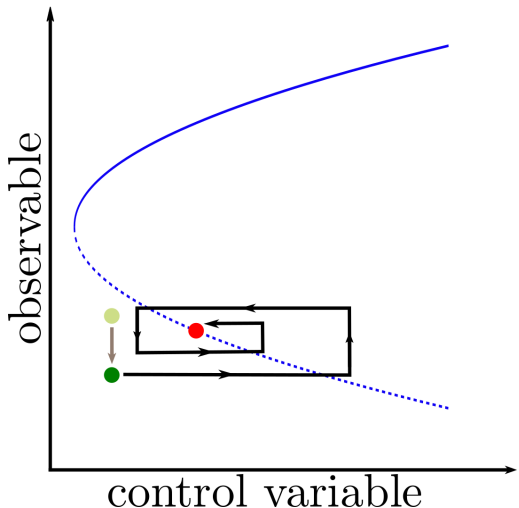}} 
\caption{Schematic dynamics of the controlled system.
	The unstable lower branch (dashed line) is curved towards smaller
	values of the observable for increasing control variable. The red (grey) dot on the lower 
	branch corresponds to an operating point.
	For an initial state below the lower branch indicated by the light green (light grey) dot, the 
	uncontrolled dynamics are such that the value of the observable 
	decreases, resulting in 
	intermediate states further below the lower branch as indicated by the green (dark grey) dot. 
	The feedback control increases the control variable until the lower branch is crossed, such 
	that the uncontrolled dynamics now result in a growing observable.
	The feedback control now decreases the value
	of the control variable until the lower branch is crossed again to enter the 
	region where the observable will decay. Iteration of this procedure will eventually
	result in convergence towards the operating point. 
}
\label{fig:control}
\end{figure}

Consider a system with two variables, a positive observable $A$ and a
control variable $R$. In fluid dynamics, $A$ could be the result of a global measurement such as the skin friction 
factor or a local measurement such as the magnitude of the turbulent fluctuations, and $R$ the Reynolds number,
which is here interpreted as a means to determine the pressure gradient as the control input.
We assume that the uncontrolled system has stationary solutions 
that appear in a saddle-node bifurcation at $(A^*, R^*)$ with an unstable lower branch $A_{LB}(R)$. The 
aim is to stabilise an operating point $(A_0,R_0)$ on the lower branch \citep{Sieber2014, Willis17}. 
Without loss of generality we further assume that the uncontrolled dynamics 
is such that the observable 
grows if it exceeds the lower branch value 
$A_{LB}(R)$,
\beq
\label{eq:dyn-free}
\dot A = \lambda (A-A_{LB}(R)) \ ,
\eeq
with $\lambda > 0$ being the Lyapunov exponent, which we assume to be independent, or a 
slowly varying function, of $R$.
To control and avoid the exponential instability, the control variable must be 
repeatedly adjusted such that the ensuing dynamics of the system results in 
convergence to the operating point, for example through an iteration procedure where 
the lower branch is crossed at each adjustment of the control variable as schematically 
illustrated in fig.~\ref{fig:control}.
For the uncontrolled dynamics as in eq.~\eqref{eq:dyn-free}, this can be achieved 
by adjusting the control variable according to
\beq
\label{eq:r-dot}
\dot{R } = - \gamma \left(R - R_0\right) - \gamma \mu (A-A_0) \ ,
\eeq
where $A_0=A(R_0)=A_{LB}(R_0)$ is the value of the observable 
at the reference point and $\gamma > 0$ and $\mu > 0$ are adjustable parameters. The signs are for the cases that $A_{LB}(R)$ decreases with $R$, i.e.
\beq
\left. \frac{d A_{LB}(R)}{dR} \right|_{R_0} = - \alpha \ ,
\eeq
with $\alpha>0$. With
$r=R-R_0$ and $a = A-A_0$ we can write
\beq
A- A_{LB}(R) = A - A_{LB}(R_0+r) \approx a + \alpha r \ ,
\eeq
so that 
eqs.~\eqref{eq:dyn-free} and \eqref{eq:r-dot} become
\beq
\label{eq:linearised-1d}
\frac{d}{dt}
\begin{pmatrix}
a \\
r
\end{pmatrix}
=
\begin{pmatrix}
\lambda  &  \lambda \alpha \\
- \gamma \mu & - \gamma
\end{pmatrix}
\begin{pmatrix}
a \\
r
\end{pmatrix} \ .
\eeq
For the operating point $(A_0,R_0)$ to be stable, the matrix on the right-hand side of 
eq.~\eqref{eq:linearised-1d} must have eigenvalues with negative real parts.
The conditions for such eigenvalues are that the trace of the matrix, as the
sum of the eigenvalues, has to be negative, and the determinant, the
product of the eigenvalues, has to be positive.
With the trace
\beq
\mbox{Tr\ }= \lambda - \gamma \ ,
\eeq
and the determinant
\beq
\mbox{det\ }= - \lambda \gamma + \lambda \gamma \alpha \mu \ ,
\eeq
the conditions for stability become
\bea
\label{eq:stability_gamma}
\gamma > \lambda\\
\label{eq:stability_mu}
\alpha\mu > 1 \ .
\eea
The conditions are such that the adjustment in $R$ (related to the
parameter $\gamma$) has to be faster than the escape
(as measured by $\lambda$). Similarly, the amplitude of the
change in the control variable with the deviation in the observable 
has to be
larger than the inverse of $\alpha$, so that the changes
in $R$ outrun the changes in $A$. For what follows it will be useful to 
visualise the stability condition \eqref{eq:stability_mu} geometrically: Since 
$\alpha$ is the slope of the tangent to the lower branch at $(A_0,R_0)$, 
the inequality \eqref{eq:stability_mu} results in a 
{\emph control line} through $(A_0,R_0)$ with a slope $1/\mu < \alpha$ which is 
shallower than that of the tangent at the operating point. 
The feedback control procedure applied by \citet{Willis17} corresponds 
in this context to an immediate adjustment of $R$, i.e. to $\gamma \to \infty$.

Before proceeding to numerical results, we briefly highlight the connection between the 
present formulation of the linear control law given in eq.~\eqref{eq:linearised-1d} and linear feedback control. 
If we combine the observable $a$ and the control variable $r$ into one state vector $\vec{x}$, then the uncontrolled 
linearised dynamics, where $r = 0$ and $\dot r = 0$, is given by 
\beq
\underbrace{\begin{pmatrix} \dot a \\ \dot r  \end{pmatrix}}_{\dot{\vec{x}}}  =   \underbrace{\begin{pmatrix} \lambda & 0 \\ 0 &  0 \end{pmatrix}}_{\mathcal{A}} \underbrace{\begin{pmatrix} a \\ r  \end{pmatrix}}_{\vec{x}} \ ,
\eeq
with Jacobian $\mathcal{A}$. The control law given in eq.~\eqref{eq:r-dot} makes $r$ time-dependent such that eq.~\eqref{eq:linearised-1d}  
can be written in classical control-theoretic form as closed-loop feedback control
\beq
\dot{\vec{x}} = \mathcal{A} {\vec{x}} - \mathcal{BK} {\vec{x}} \ ,
\eeq
where $\mathcal{B}$ is the control matrix and, for stabilisation according to linear optimal control or full state feedback, 
the matrix $\mathcal{K}$ must be chosen such that 
$ \mathcal{A} - \mathcal{BK} = \begin{pmatrix} \lambda  &  \lambda \alpha \\ - \gamma \mu & - \gamma \end{pmatrix}$ 
has only eigenvalues with negative real part, see  e.~g.~\citet{Anderson90,Sontag98,Burl99}.  

\subsection{Two-dimensional linear model}
\label{sec:1d-model}
\begin{figure}
	{\includegraphics[width=0.48\columnwidth]{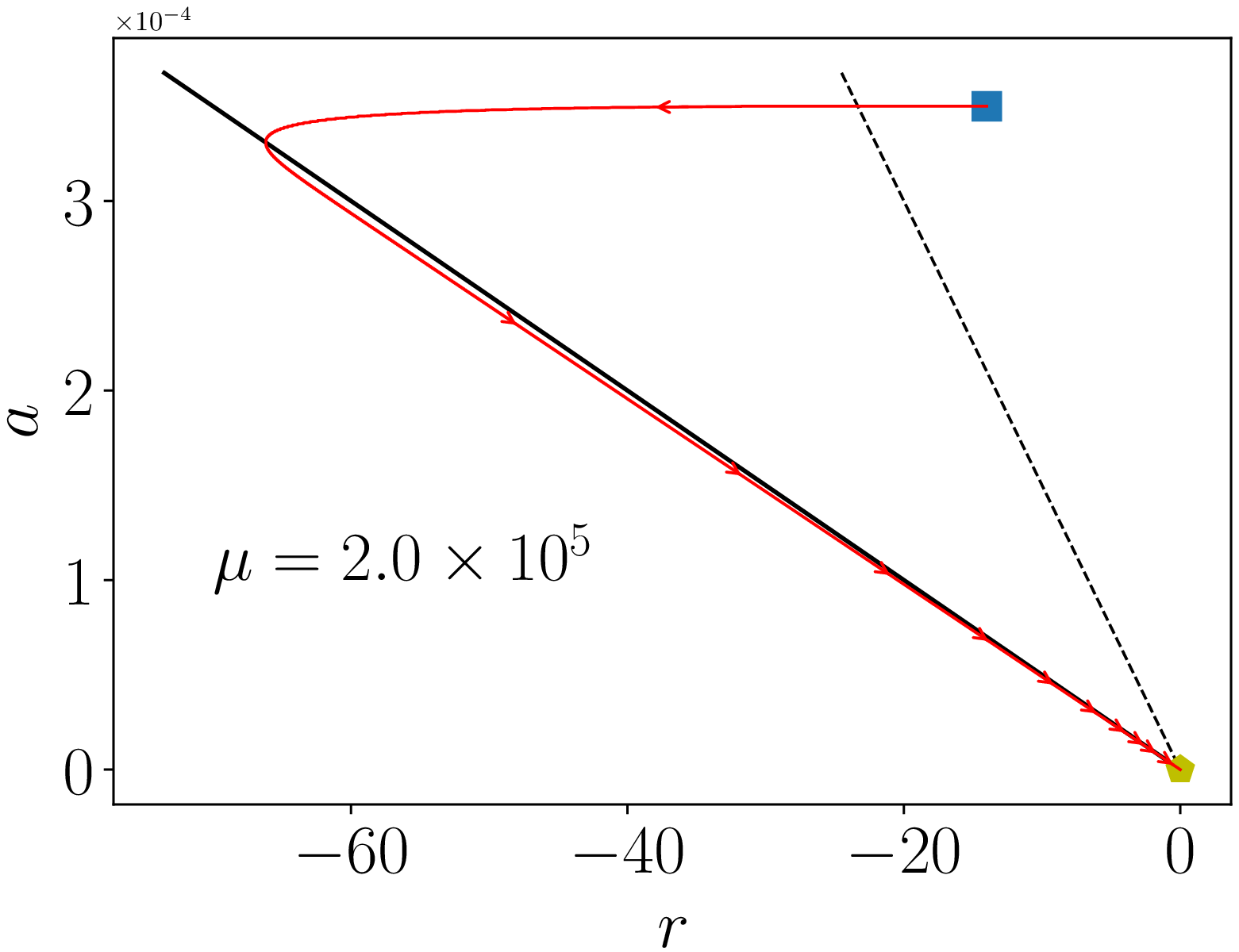}} 
	{\includegraphics[width=0.48\columnwidth]{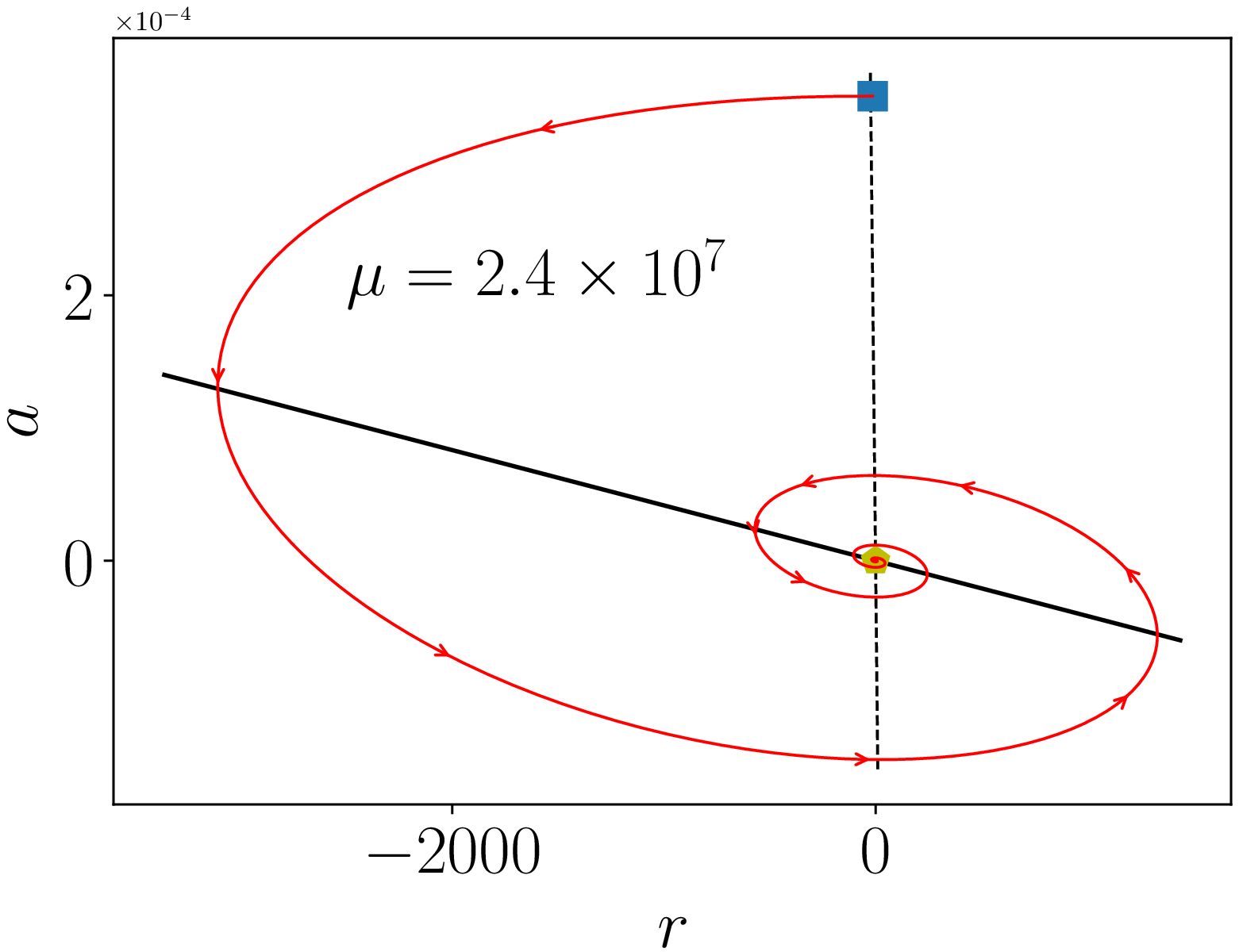}} 
\caption{
Stabilisation of the linear model system given by eq.~\eqref{eq:linearised-1d}. 
	The (linear) lower branch is indicated by the dashed line, it crosses the control line (solid black) at the 
	operating point. The time evolution of the system follows the red (grey) curve starting at the blue (dark grey) square.
	Left: monotonic relaxation for $\mu = 2 \times 10^5$ corresponding to negative real eigenvalues. 
	Right: oscillatory relaxation for $\mu = 2.4 \times 10^7$ corresponding to complex eigenvalues with negative real parts. 
}
\label{fig:1D-model}
\end{figure}

Before applying the feedback control to a high-dimensional dynamical system
such as channel flow, we consider the dynamics of the linearised
two-dimensional (2D)  
system given by eq.~\eqref{eq:linearised-1d}, with Lyapunov
exponent $\lambda = 0.01$, and lower branch slope $\alpha = 1.5 \times 10^{-5}$. These values
correspond to measurements of $\alpha$ and $\lambda$ for an edge state in DNS
of channel flow, which will be discussed in further detail in
sec.~\ref{sec:numerics}.  Figure \ref{fig:1D-model} presents phase-space trajectories 
of this system for $\gamma = 1$ and two different values of the control
strength $\mu$, i.e. $\mu = 2 \times 10^5$ and $\mu = 2.4 \times 10^7$. The
tangent line as indicated in orange (light grey) has a steeper slope than the
control line (blue/dark grey) in both cases, as required by
eq.~\eqref{eq:stability_mu}, and both lines cross at the operating point. The
time evolution follows the green/grey curve, beginning at the red/grey points
located in the top right quadrants of the two panels, and it ends at the
operating point. That is, in both cases the operating point has been stabilised.

In both cases the instability has been removed, leading to
eigenvalues of the matrix in eq.~\eqref{eq:linearised-1d} that have negative
real parts. The eigenvalues do not only yield information on the stability of
an equilibrium in the controlled system, they also determine the dynamic
relaxation process. For real eigenvalues we expect monotonic exponential
relaxation, while complex eigenvalues with non-zero imaginary part lead to an
oscillatory approach to the stabilised equilibrium.  In the present linear 
2D
model system, the eigenvalues are real for $\mu = 2 \times 10^5$ and complex
for $\mu = 2.4 \times 10^7$, and the relaxation towards the equilibrium does
indeed proceed differently for the two values of the control strength.  For
$\mu = 2 \times 10^5$ the relaxation proceeds monotonically along the control
line as shown in the left panel of fig.~\ref{fig:1D-model}, while $\mu =
2.4 \times 10^7$ results in oscillatory relaxation as shown in the right panel
of fig.~\ref{fig:1D-model}.  The latter is reminiscent of the schematic
behaviour illustrated in fig.~\ref{fig:control}. 

\subsection{Effect on the stable directions}
\label{sec:stable-dir}
\begin{figure}
	\centerline{\includegraphics[width=0.6\columnwidth]{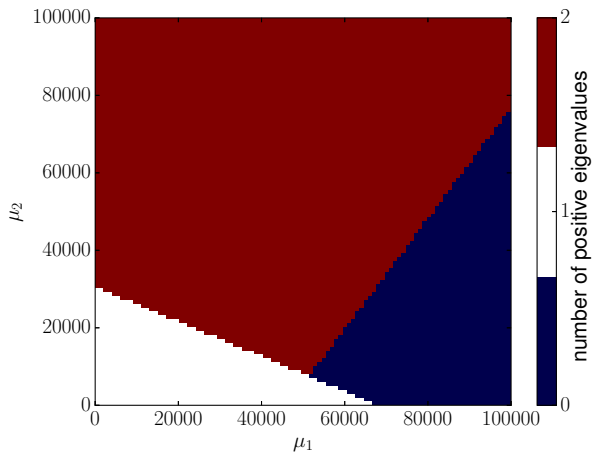}} 
\caption{
	Destabilisation of a stable direction for the 3D-model system given by eq.~\eqref{eq:linearised-2d}.
	The colour coding represents the number of eigenvalues of the matrix on the right-hand side 
	of eq.~\eqref{eq:linearised-2d} that have positive real parts. The coupling of the control to the 
	unstable and stable directions is parametrised by $\mu_1$ and $\mu_2$, respectively. 
}
\label{fig:mu1mu2}
\end{figure}

Equilibria in higher-dimensional systems can have several stable and unstable
directions. Even if we assume that only one direction is unstable, as is
generally the case for edge states in canonical wall-bounded parallel shear
flows, a one-dimensional control procedure may not only have the desired influence on the
unstable direction, it may also couple to the stable directions. 
This effect is known in control theory, where its mitigation is essential
in the design of successful controllers \citep{Barbagallo09}. 
In order to
illustrate what the consequences of such a coupling can be, we consider a
three-dimensional (3D) 
extension of the 2D-model given in linearised
form in eq.~\eqref{eq:linearised-1d}
\beq
\label{eq:linearised-2d}
\frac{d}{dt}
\left(
\begin{matrix}
 r \\
 a_1 \\
 a_2 \\
\end{matrix}
\right)
=
\left(
\begin{matrix}
        - \gamma & - \gamma \mu_1 & -\gamma \mu_2 \\
        \lambda_1 \alpha_1  &  \lambda_1  & 0 \\
        -\lambda_2 \alpha_2  &  0  & -\lambda_2
\end{matrix}
\right)
\left(
\begin{matrix}
 r   \\
 a_1 \\
 a_2 \\
\end{matrix}
\right) \ ,
\eeq
where $a_1$ corresponds to the unstable and $a_2$ to the stable direction with
$\lambda_1 > 0$ and $\lambda_2 > 0$. The dynamics are coupled to the control
procedure through $\mu_1$ and $\mu_2$, respectively.
For simplicity, we assume that the stable and
unstable directions decouple.  
For $a_1 = a$, $\lambda_1 = \lambda$ and $\alpha_1 = \alpha$ as in
fig.~\ref{fig:1D-model}, we construct arbitrary stable directions by randomly choosing $a_2 > 0$, $\lambda_2 > 0$ and $\alpha_2 > 0$
to avoid a specific configuration. Subsequently and for fixed values of 
$a_2 > 0$, $\lambda_2 > 0$ and $\alpha_2 > 0$, we 
calculate the number of eigenvalues of the matrix on the right-hand side of
eq.~\eqref{eq:linearised-2d} that have a positive real part as a function of
$\mu_1$ and $\mu_2$. 
An example of the results obtained from such a
calculation is shown in fig.~\ref{fig:mu1mu2}. If the control is  weakly
coupled to the dynamical system, we find one eigenvalue with positive real
part, as expected for a system with one stable and one unstable direction.
Increasing $\mu_1$ for small $\mu_2$ eventually stabilises the operating point, which
can also be expected from the results in the 1D case. However, we find a large
part of parameter space where one or two eigenvalues have a positive real part, hence
the control is not able to stabilise the operating point if it overlaps
significantly with the stable direction.  

In summary, the success of the control strategy in higher-dimensional systems
depends on how the dynamics along the stable directions couple to the control. Stabilisation of
the operating point then requires  a control strategy that acts on a hyperplane
orthogonal to all stable directions.  Such a strategy can be constructed in
numerical simulations only, and we will come back to this point in 
sec.~\ref{sec:revised-control}. In practice, the control is more likely to
destabilise stable directions with a small negative real part,  
which suggests that it may be sufficient to design the control to be orthogonal 
to the least stable directions in order to achieve stabilisation. 
Similar procedures are indeed sometimes applied in control theory in 
the context of model reduction \citep{Akervik07} and will 
be successful provided the chosen modes are observable and controllable
\citep{Barbagallo09}.


\section{Linear feedback control for the Navier-Stokes equations}
\label{sec:control_nse}
Having introduced a general one-dimensional feedback control strategy and 
discussed its properties in low-dimensional model systems, we
now turn to its application to wall-bounded shear flows, whose dynamics
is governed by the incompressible Navier\textendash Stokes equations
\begin{align}
	\label{eq:Navier-Stokes}
	\partial_t \vec{u} + \vec{u} \cdot \nabla \vec{u} & = 
	-\nabla p + \nu\Delta \vec{u} + \frac{1}{\rho}\vec{f}\ , \\
        \nabla \cdot \vec{u} & = 0 \ , 
\end{align}
where $\vec{u}$ is the velocity field, $p$ the pressure divided by the constant
density $\rho$, $\nu$ the kinematic viscosity and $\vec{f}$ a force that drives
the flow. 
The implementation of the feedback procedure introduced in sec.~\ref{sec:theory}
requires a choice of observable and control variable. Here, care must be taken
in the non-dimensionalisation of eq.~\eqref{eq:Navier-Stokes}, as the choice of
control variable may result in the usual characteristic scales becoming
time-dependent.  Furthermore, eq.~\eqref{eq:Navier-Stokes} must be supplemented
with an auxiliary equation that describes the time evolution of the control
variable as a function of the observable. The feedback loop is then closed by
coupling the control variable to eq.~\eqref{eq:Navier-Stokes}. 

In principle, there are two conceptual choices for the control variable, one
that results in a modulation of the flow and one that results in an adjustment
of $\vec{f}$.  Since eq.~\eqref{eq:Navier-Stokes} is usually made dimensionless
using a characteristic length scale $h$ and a reference  velocity $U_0$, the
choice of control variable must be such that $U_0$ and $h$ remain
time-independent.  Otherwise the dimensionless form of
eq.~\eqref{eq:Navier-Stokes} is not applicable any longer because the
time-derivative does not commute with the now time-dependent reference velocity
$U_0(t)$. This occurs if the feedback is implemented through a modulation of
the flow. Therefore, we focus here on the second possibility, that of an
adjustment of $\vec{f}$ in response to a control variable. Assuming that
$\vec{f}(t)$ fluctuates around a reference state $\vec{f}_0$, the velocity
scale $U_0$ that is associated with that particular value of the force is used
to rescale eq.~\eqref{eq:Navier-Stokes}.  Specifically, the forcing is made
dimensionless in units of $h/U_0^2$, and variations in the force can be
measured in the same units.  In what follows, $U_0$ is the laminar centerline
velocity and $h$ the half-height of the channel.

\subsection{A pressure-based control strategy}
\label{sec:strategy}
For pressure-driven pipe or channel flow, the control input $\vec{f}$ 
can be identified with a time-dependent streamwise pressure gradient 
$dP/dx(t)\vec{e}_x$ that fluctuates around a reference value $(dP/dx)_0 \vec{e}_x$. 
The controlled system in non-dimensionalised form then reads 
\begin{align}
	\label{eq:nse-dPdx1}
	\partial_t \vec{u} + \vec{u} \cdot \nabla \vec{u} +\nabla p & - \frac{1}{\Rey}\Delta \vec{u} + \left(\frac{dP}{dx}\right)_0 \vec{e}_x = -\frac{dP}{dx}(t)\vec{e}_x \ , \\
	\label{eq:nse-dPdx2}
        \nabla \cdot \vec{u} & = 0 \ ,
\end{align}
\begin{align}
	\label{eq:nse-dPdx3}
        \dot{R}  & = - \gamma \left(R - R_0\right) - \gamma \mu (A-A_0) \ , \\
	\label{eq:nse-dPdx4}
	\frac{dP}{dx}(t) &= -\frac{2}{R_0}\left(\frac{R(t)}{R_0}-1\right) \ ,
\end{align}
with $A$ being an observable, $R$ the control variable with $(R_0, A_0)$
defining the operating point. The last equation, which implements the feedback,
is based on $R$ representing a Reynolds number such that $R = R_0 = Re$ 
results in no control input and the reference pressure gradient is recovered.
The time-dependent Reynolds number
that is used by \citet{Willis17} cannot be realised with a change in the
pressure gradient or similar, since that would give a different velocity scale,
as discussed above.  As it stands, a modulation in Reynolds number can only be
obtained as a consequence of variations in viscosity, which is difficult to
achieve in experiments. 

In order to stabilise the operating point, the control must overlap with the 
expanding directions of the operating point's tangent space. Since the linear 
operator representing the linearised Navier\textendash Stokes dynamics close to the operating 
point is non-normal, its eigenvectors are not orthogonal. That is, it is 
in principle possible to stabilise an operating point with a one-dimensional
control procedure, provided that all unstable directions overlap.
Here, the proposed feedback control acts in the streamwise direction only and 
it is translationally invariant in both streamwise and spanwise directions. 
That is, it can only stabilise unstable directions that have a streamwise
component with a non-zero streamwise and spanwise mean.  
Periodic instabilities, for instance, cannot be stabilised. 
This is an example of a more general effect that symmetries, 
translational invariance being an example thereof, have on
controllability and observability in linear feedback control \citep{Grigoriev2000}.
Formally speaking, an $n$-dimensional system is controllable if the vectors $w_k^l = \mathcal{A}^{n-l}b_k$ for $1 \leq k \leq n$ and $1 \leq l \leq m$, 
where $\mathcal{A}$ is the Jacobian governing the linearised dynamics  
and $b_k$ denotes the $k^{th}$ column vector of the control matrix $\mathcal{B} = (b_1, \hdots, b_m)$, form a basis of the 
tangent space at the operating point. An equivalent formulation is that 
each eigenmode must have nonzero overlap with at least one column vector of $\mathcal{B}$. 
Symmetries may lead to eigenspaces of the linear operator $\mathcal{A}$ 
of dimension larger than one, and hence basis vectors of such eigenspaces 
exist which are orthogonal to all $b_k$ \citep{Grigoriev98,Grigoriev2000}.
Similar issues concern in principle also the question of observability, however, 
such complications do not arise in the present context 
as we have access to the full state of the system at any point in time.

\subsection{Adjoint-based control}
\label{sec:adjoint_control}
The potentially destabilising effect of the control given by
eqs.~\eqref{eq:nse-dPdx1}-\eqref{eq:nse-dPdx4} calls for a 
strategy that acts on the unstable direction only.  
In what follows we
construct a control that acts on a hyperplane orthogonal to the stable subspace
of the ECS's tangent space and hence cannot couple and destabilise the
contracting directions.  
Similar approaches are used in controlling linear, infinite-horizon
problems. There, the optimal control strategy
is of feedback type and proceeds by projection of the state vector 
onto its unstable eigenspace and an appropriate choice of 
coupling coefficients such that 
the linear operator representing the controlled system has only stable eigenmodes 
\citep{Anderson90,Burl99}.

We consider a general $n$-dimensional
dynamical system
\beq
\label{eq:dynamical_system}
\dot \xi = F(\xi)\ , 
\eeq
where $F$ is a differentiable function that governs the time-evolution of
$\xi$.  In the present application $\xi$ represents the Galerkin-truncated
velocity field and $F$ the time evolution given by the appropriately truncated
version of eq.~\eqref{eq:Navier-Stokes} in terms of a finite number of coupled
ordinary differential equations. Let $\xi_0$ correspond to the operating point,
then the linearised dynamics close to $\xi_0$ are given by
\beq
\dot \delta \xi = J_F\delta \xi \ , 
\eeq
where $J_F=J_F(\xi_0)$ is the Jacobian of $F$ at $\xi_0$.  The tangent space at
$\xi_0$ is then spanned by the right eigenvectors $\{v_i\}_{(i = 1, \hdots,
n)}$ of $J_F$.  Since $J_F$ is in general non-normal, the $\{v_i\}$ are not
mutually orthogonal, i.e. $(v_i,v_j) \neq \delta_{ij}$, with $(\cdot, \cdot)$
being an inner product on the tangent space at $\xi_0$. Hence a control
procedure that overlaps with the unstable directions may also overlap with the
stable and the marginal ones. However, the dual basis $\{v_i^*\}$,
defined as the set of linear maps from the tangent space at $\xi_0$ to 
$\mathbb{C}$ satisfying $v_i^*(v_j) = \delta_{ij}$ 
satisfies the desired
bi-orthogonality constraints by definition $(v_i,v_j^*) := v_i^*(v_j) = \delta_{ij}$. If we
have $k<n$ unstable directions, $v_1,..,v_k$, a control that is constructed as
a linear combination of the duals $v_1^*,..,v_k^*$ will be orthogonal to all
stable and marginal directions.  More specifically, the purpose of a feedback control with control input
$f(\xi)$ is to stabilise $\xi_0$, that is, to ensure that all eigenvalues of
$J_{F} + J_f$, 
where $J_f=J_f(\xi_0)$ is the Jacobian of $f$ at $\xi_0$,  
have negative or zero real parts.  For reasons of clarity and conciseness, we assume from now on that
$J_{F}$ has one expanding direction $v_e$, as the generalisation to more unstable directions is straightforward. 
If we construct $f$ to act along $v_e^*$ such that the controlled dynamical system is given by
\beq
\label{eq:ctrl-dynamical_system}
\dot \xi = F(\xi) + f(\xi) = F(\xi) + \kappa(\xi)v_e^* \ , 
\eeq
where $\kappa$ is a function of $\xi$ implementing the feedback, 
then the controlled linearised system is
\beq
\label{eq:ctrl-linearised}
\dot \delta \xi = \left(J_F + J_f \right)\delta \xi =  \left(J_F + v_e \otimes \nabla\kappa \right)\delta \xi \ , 
\eeq
where $\nabla \kappa$ denotes the gradient of $\kappa$ at $\xi_0$ and we use tensor product notation 
for $J_f$, that is $(a \otimes b)_{ij} := a_i b_j$ for two generic vectors $a$ and $b$.  Since the
dimension of the tangent space at any point equals that of the underlying
manifold, we can expand $\delta \xi$  at any point in time in terms of the
basis $v_i$
\beq
\delta \xi(t) = \sum_i a_i(t)v_i \ ,  
\eeq
where $a_i$ are time-dependent coefficients. Equation \eqref{eq:ctrl-linearised}
becomes
\begin{align}
\sum_i \dot a_i(t)v_i & = \left(J_F + v_e \otimes \nabla\kappa \right) \sum_i a_i(t)v_i  
		       = \sum_{i} \left(\lambda_i+ \sum_jk_j v_e \otimes v_j \right)a_i(t)v_i  \nonumber \\ 
		      & = \sum_{i} \lambda_i a_i(t)v_i + \sum_{i,j}a_i(t)k_j (v_j^*, v_i) v_e  \nonumber \\ 
		      & = \sum_{i} \lambda_i a_i(t)v_i + \sum_{i,j}a_i(t)k_j \delta_{ij} v_e   \nonumber \\ 
		      & = \sum_{i \neq e} \lambda_i a_i(t)v_i 
			+ \left(\lambda_e a_e(t)+ \left(\sum_i k_i a_i(t) \right) \right)v_e  \ , 
\end{align}
where $\lambda_i$ are the eigenvalues of $J_F$ and $k_j = (\nabla \kappa^*,
v_j)$.  By taking the inner product of both sides of this equation with $v_e^*$
it can be seen that $\xi_0$ is stabilised if $k_i = 0$ for $i \neq e$ and if 
\beq
\lambda_e + k_e  \leqslant 0 \ ,
\eeq
that is, the gradient of the feedback function $\kappa$ at the operating point
must be colinear with the unstable direction.  In the present example of
channel flow, the control input $\kappa$ is determined by the choice of observable. An observable
that is quadratic in the velocity field will result in $\nabla \kappa$ being
colinear with the operating point. If the latter
then has a significant overlap with the unstable direction, the choice of
observable may work well. Close inspection of the unstable direction can yield
further information, for example if the instability is mostly in span- or
wall-normal directions, the cross-flow energy is a good observable.

For time-independent operating points, i.e. equilibria of
eq.~\eqref{eq:dynamical_system}, with one unstable eigenmode, the
implementation of such a control procedure results in replacing the unit vector
$\vec{e}_x$ on the right-hand side of eqs.~\eqref{eq:nse-dPdx1} the dual
of the solution's unstable eigenmode, $v^*_e$, which has been normalised to be a unit vector. 
The generalisation to more unstable
directions is straightforward.  For travelling wave or periodic solutions, the
implementation is slightly more complicated as the time-dependence of the
target state has to be accounted for.  For a wave travelling in streamwise
direction with speed $c$ the adjoint-based control strategy is given by
eqs.~\eqref{eq:nse-dPdx2}-\eqref{eq:nse-dPdx3}, with \eqref{eq:nse-dPdx1}
replaced by 
\beq
\label{eq:nse-dPdx1-dd}
\partial_t \vec{u} + \vec{u} \cdot \nabla \vec{u} 
                   +\nabla p - \frac{1}{\Rey}\Delta \vec{u} 
		   + \left(\frac{dP}{dx}\right)_0 \vec{e}_x 
		   = \frac{2}{R_0}\left(\frac{R(t)}{R_0}-1\right)\sigma_c(t)(v^*_e) \ . 
\eeq
where $\sigma_c$ is the shift operator in streamwise direction
\beq
\sigma_c(t): \vec{u}(x,y,z) \mapsto \vec{u}(x+ct,y,z) \ . 
\eeq
Shifts in spanwise direction can be accounted for analogously. 

Projections onto bi-orthogonal bases, stable and unstable eigenmodes used
in the feedback strategy proposed here being only one example thereof, are used
in controlling high-dimensional systems where the algorithm requires a
reduction of the number of degrees of freedom to become viable
\citep{Antoulas01,Lauga03,Lauga04,Akervik07,Ehrenstein08,Henningson08,
Barbagallo09}.  There, a high-dimensional system is modelled by projection onto
a lower dimensional subspace spanned by an appropriately chosen set of basis
modes, and a control strategy for the reduced system is calculated.  In order
for this control strategy to work on the full system, the subspace must, of
course, include all unstable eigenmodes, but more importantly also the set of
stable eigenmodes that are triggered by the control.  \citet{Ehrenstein08}
successfully stabilised an unstable flow by projection onto a subset of stable
eigenmodes, however, this is not a strategy that works generically, and
sometimes other bases such as proper orthogonal decomposition (POD) modes
constitute a better choice \blue{\citep{Rowley2005,Rowley2017}}. 

We point out that the method defined in Eq.~\eqref{eq:nse-dPdx1-dd} is 
in general not experimentally applicable. First, it requires information on the 
invariant solution and its stable and unstable directions, which is usually 
not attainable in experiments. Second, the applied forcing cannot 
be realised in practise, as it will need to act on the entire flow field and at all scales.
Here, we introduce this method as a simple and clear means to 
discuss the limitations of global one-dimensional feedback control in general and to specifically 
emphasise
(i) what in principle needs to be done in order to stabilise an invariant solution,
(ii) what difficulties arise, in particular concerning the choice of observable,
(iii) which obstacles need to be overcome when 
considering to devise feedback control methods                   
aimed at finding and continuing invariant solutions in parameter space.
Before proceeding to use these methods to stabilise simple invariant solutions 
and a subsequent discussion of general issues concerning the application of 
linear feedback control in this context, we briefly outline the numerical 
method and \blue{then describe} the target states.

\section{Datasets and numerical details}
\label{sec:numerics}
Direct numerical simulations (DNS) of channel flow have been carried out using
the pseudospectral open-source code {\tt channelflow2.0}
\citep{Gibson2014,chflow18}.  The code numerically solves the incompressible
Navier\textendash Stokes equations  \eqref{eq:Navier-Stokes} in a rectangular domain with
periodic boundary conditions in streamwise and spanwise $(x,z)$ directions, and
no-slip boundary conditions in the wall-normal ($y$) direction.  The spatial
discretisation is obtained through Fourier expansions in $x$- and
$z$-directions using $N_x$ and $N_z$ collocation points, respectively, and a
Chebyshev expansion in the $y$-direction on $N_y$ points.  Aliasing errors in
the periodic directions are removed by $2/3$-Galerkin truncation
\citep{Orszag71}. A third-order semi-implicit Adams-Bashforth scheme is used
for the temporal discretisation.  The code has been adapted to run the
controlled simulations as the core dynamical system in order to make use of the
methods for numerical stability analysis provided in {\tt channelflow2.0}. 
As discussed in the Introduction, the aim here is to stabilise the simplest
invariant solutions with one unstable direction, that is, travelling-wave type 
edge states in minimal flow units.
For this reason all simulations in this study are
carried out in a short computational domains of size $L_x/h \times L_y/h \times
L_z/h = 2 \pi \times 2 \times 2\pi$ and $L_x/h \times L_y/h \times L_z/h = 2
\pi \times 2 \times \pi$.  
Further details of all simulations are summarised in
table \ref{tab:simulations}.

The construction of the adjoint feedback procedure requires access to the 
stable, neutral and unstable subspaces of the uncontrolled system. The corresponding eigenmodes 
of the Jacobian of the uncontrolled system 
were calculated by Arnoldi iteration and marginal and stable eigenmodes were 
subsequently used to construct the dual basis. 
Stability analyses of the pressure-controlled system were also carried out using the Arnoldi method.

\begin{table}
  \begin{center}
\def~{\hphantom{0}}
  \begin{tabular}{lccccccccccc}
	  id  & $L_x/h$ & $L_y/h$ & $L_z/h$ & $N_x$ & $N_y$ & $N_z$ & control type & observable & $\Rey_0$ & $\mu$ & $\|\delta \vec{u} \|_2 / \|\vec{u}^*\|_2$ \\[3pt]
	  TW1-A1    & $2\pi$ & 2 & $2\pi$ & 32 & 49 & 48 & $dP/dx$ & $L_2$-norm & 1395 & $2 \times 10^5$ & 0.11\\
	  TW1-A2    & $2\pi$ & 2 & $2\pi$ & 32 & 49 & 48 & $dP/dx$ & $L_2$-norm & 1395 & $6 \times 10^5$ & 0.11\\
	  TW1-A3    & $2\pi$ & 2 & $2\pi$ & 32 & 49 & 48 & $dP/dx$ & $L_2$-norm & 1395 & $10^6$          & 0.11\\
	  TW1-A-stab& $2\pi$ & 2 & $2\pi$ & 32 & 49 & 48 & $dP/dx$ & $L_2$-norm & 1395 & $0-10^6$        & $4 \times 10^{-5}$ \\
	  TW1-B1    & $2\pi$ & 2 & $2\pi$ & 32 & 49 & 48 & $dP/dx$ & $C_f$      & 1395 & $2 \times 10^5$ & 0.11\\
	  TW1-B2    & $2\pi$ & 2 & $2\pi$ & 32 & 49 & 48 & $dP/dx$ & $C_f$      & 1395 & $10^6$          & 0.11\\
	  TW1-B3    & $2\pi$ & 2 & $2\pi$ & 32 & 49 & 48 & $dP/dx$ & $C_f$      & 1395 & $3 \times 10^6$ & 0.11\\
	  TW1-C1    & $2\pi$ & 2 & $2\pi$ & 32 & 49 & 48 & $v_e^*$ & $L_2$-norm & 1395 & $10^6$    & 0.11\\
	  TW1-C2    & $2\pi$ & 2 & $2\pi$ & 32 & 49 & 48 & $v_e^*$ & $L_2$-norm & 1395 & $10^7$    & 0.11\\
	  TW1-D1    & $2\pi$ & 2 & $2\pi$ & 32 & 49 & 48 & $v_e^*$ & $Ecf$      & 1395 & $10^9$    & 0.11\\
	  TW1-D2    & $2\pi$ & 2 & $2\pi$ & 32 & 49 & 48 & $v_e^*$ & $Ecf$      & 1395 & $10^{10}$ & 0.11 \\
	  TW-sym-A1 & $2\pi$ & 2 & $ \pi$ & 48 & 65 & 48 & $dP/dx$ & $L_2$-norm & 1010  & $10^4$  & 0.06 \\
	  TW-sym-A2 & $2\pi$ & 2 & $ \pi$ & 48 & 65 & 48 & $dP/dx$ & $L_2$-norm & 1010  & $5 \times 10^4$  & 0.06 \\
	  TW-sym-A3 & $2\pi$ & 2 & $ \pi$ & 48 & 65 & 48 & $dP/dx$ & $L_2$-norm & 1010  & $10^5$  & 0.06 \\
	  TW-sym-B1 & $2\pi$ & 2 & $ \pi$ & 48 & 65 & 48 & $dP/dx$ & $Ecf$      & 1010  & $3 \times 10^5$  & 0.06 \\
	  TW-sym-B2 & $2\pi$ & 2 & $ \pi$ & 48 & 65 & 48 & $dP/dx$ & $Ecf$      & 1010  & $5 \times 10^5$ & 0.06 \\
	  TW-sym-B3 & $2\pi$ & 2 & $ \pi$ & 48 & 65 & 48 & $dP/dx$ & $Ecf$      & 1010  & $6 \times 10^5$ & 0.06 \\
	  TW-sym-C1 & $2\pi$ & 2 & $ \pi$ & 48 & 65 & 48 & $v_e^*$ & $L_2$-norm  & 1010  & $3 \times 10^5$  & 0.06 \\
	  TW-sym-C2 & $2\pi$ & 2 & $ \pi$ & 48 & 65 & 48 & $v_e^*$ & $L_2$-norm  & 1010  & $3.5 \times 10^5$  & 0.06 \\
	  TW-sym-C3 & $2\pi$ & 2 & $ \pi$ & 48 & 65 & 48 & $v_e^*$ & $L_2$-norm  & 1010  & $4.75 \times 10^5$ & 0.06 \\
	  TW-sym-C4 & $2\pi$ & 2 & $ \pi$ & 48 & 65 & 48 & $v_e^*$ & $L_2$-norm  & 1010  & $5.25 \times 10^5$  & 0.06 \\
	  TW-sym-C5 & $2\pi$ & 2 & $ \pi$ & 48 & 65 & 48 & $v_e^*$ & $L_2$-norm  & 1010  & $6 \times 10^5$  & 0.06 \\
  \end{tabular}
  \caption{
	   Simulation parameters and observables. The Reynolds number is  
	   $\Rey_0 = U_0 h/\nu$, where $U_0$ is the laminar centerline velocity,
	   $h = L_y/2$ half the domain height, $\nu$ the kinematic viscosity,
	   $\mu$ the control strength as in eq.\eqref{eq:nse-dPdx3}
	   and $\delta \vec{u}$ the perturbation about the respective operating point $\vec{u}^*$. 
	   The adjustment rate in eq.\eqref{eq:nse-dPdx3} is $\gamma = 1$ in all cases. 
	   The control type $dP/dx$ refers to the pressure-based control given in 
	   eqs.\eqref{eq:nse-dPdx1}-\eqref{eq:nse-dPdx4}, while that labelled $v_e^*$ 
	   refers to the control along the dual vector of the unstable direction 
	   implemented according to eqs.\eqref{eq:nse-dPdx3}, \eqref{eq:nse-dPdx4} and \eqref{eq:nse-dPdx1-dd}. 
	   The number of Fourier modes in $x$ and $z$-directions, $N_x$ and $N_z$, contain the dealiased modes.  
	  }
  \label{tab:simulations}
  \end{center}
\end{table}

\subsection{Operating points: Travelling waves in channel flow}
\label{sec:operating_points}

The invariant solutions
we wish to stabilise are travelling waves with one unstable
direction, they are edge states in minimal flow units, 
which have been obtained by means of edge tracking 
in simulations with constant pressure gradient. 
In general, constant-flux simulations with variable pressure gradient 
are closer to experimental conditions, especially in small domains. For 
travelling wave solutions this issue is mitigated as they are relative fixed points and 
thus have no dynamics. As such, travelling wave solutions obtained with the constant-flux constraint 
result in a constant pressure gradient.  

\begin{figure}
	\centering
	{\includegraphics[width=0.8\columnwidth]{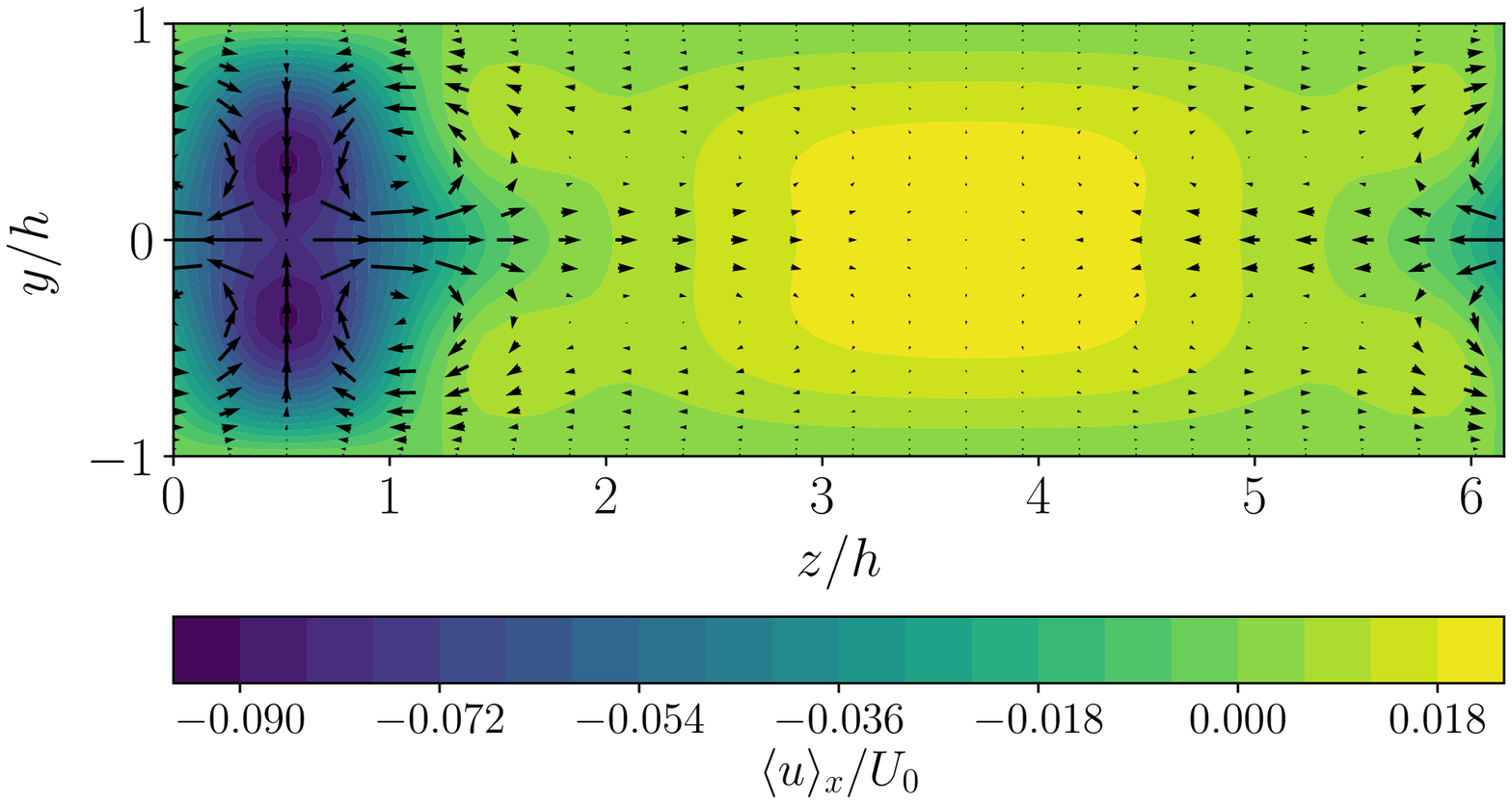}} 
	{\includegraphics[width=0.8\columnwidth]{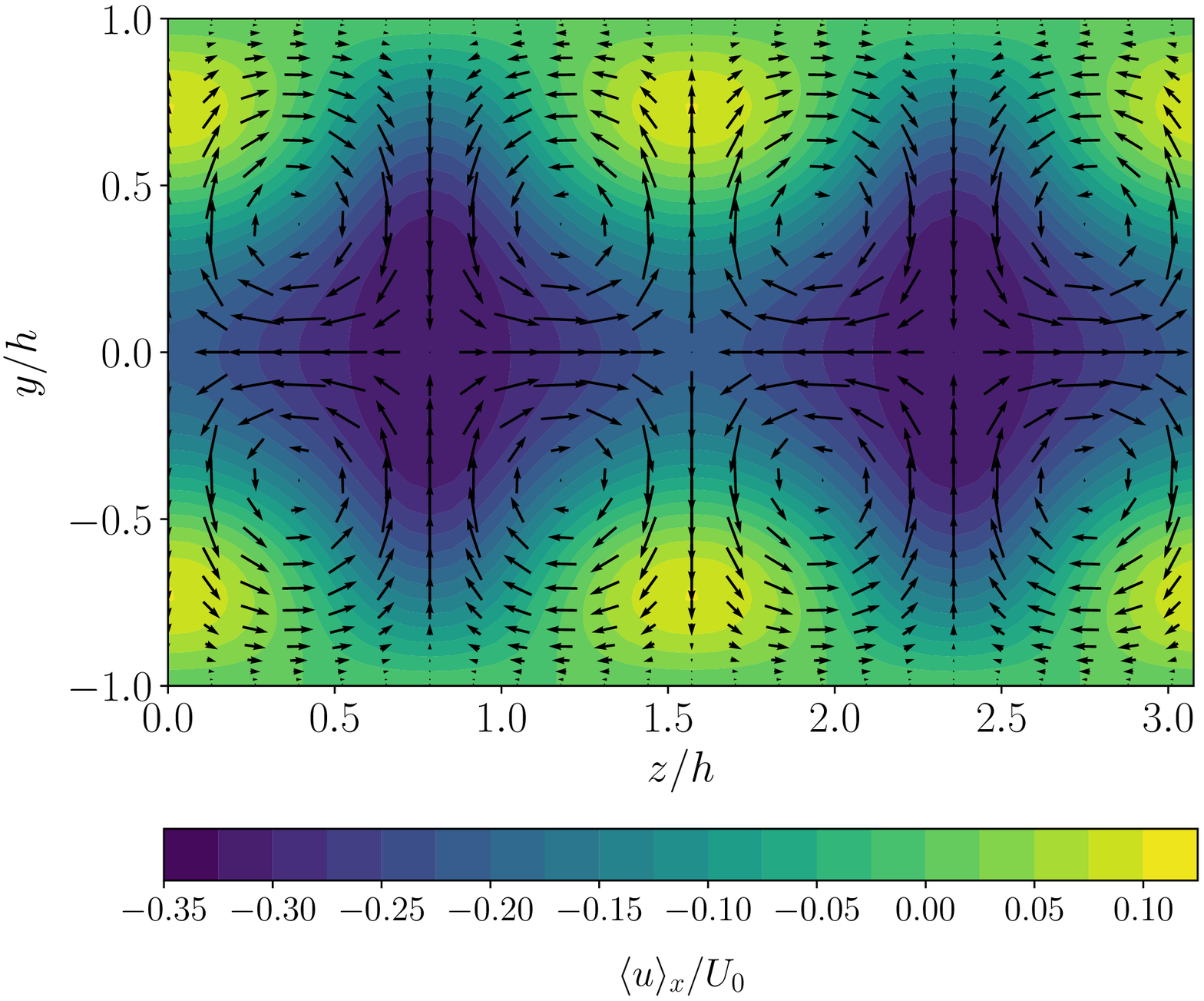}} 
\caption{
	Visualisation of the edge states showing the deviation of the streamwise average of 
	the streamwise velocity component, $\langle u \rangle_x$ from the laminar profile.
	The crossflow $(v,w)$ is indicated by the superimposed arrows.
	Top: edge state at $Re_0 = 1394$, bottom: edge state at $Re_0 = 1010$ calculated in 
	its \blue{symmetry-invariant} subspace. 
}
\label{fig:visu-edge}
\end{figure}

\begin{figure}
	\centering
	{\includegraphics[width=0.8\columnwidth]{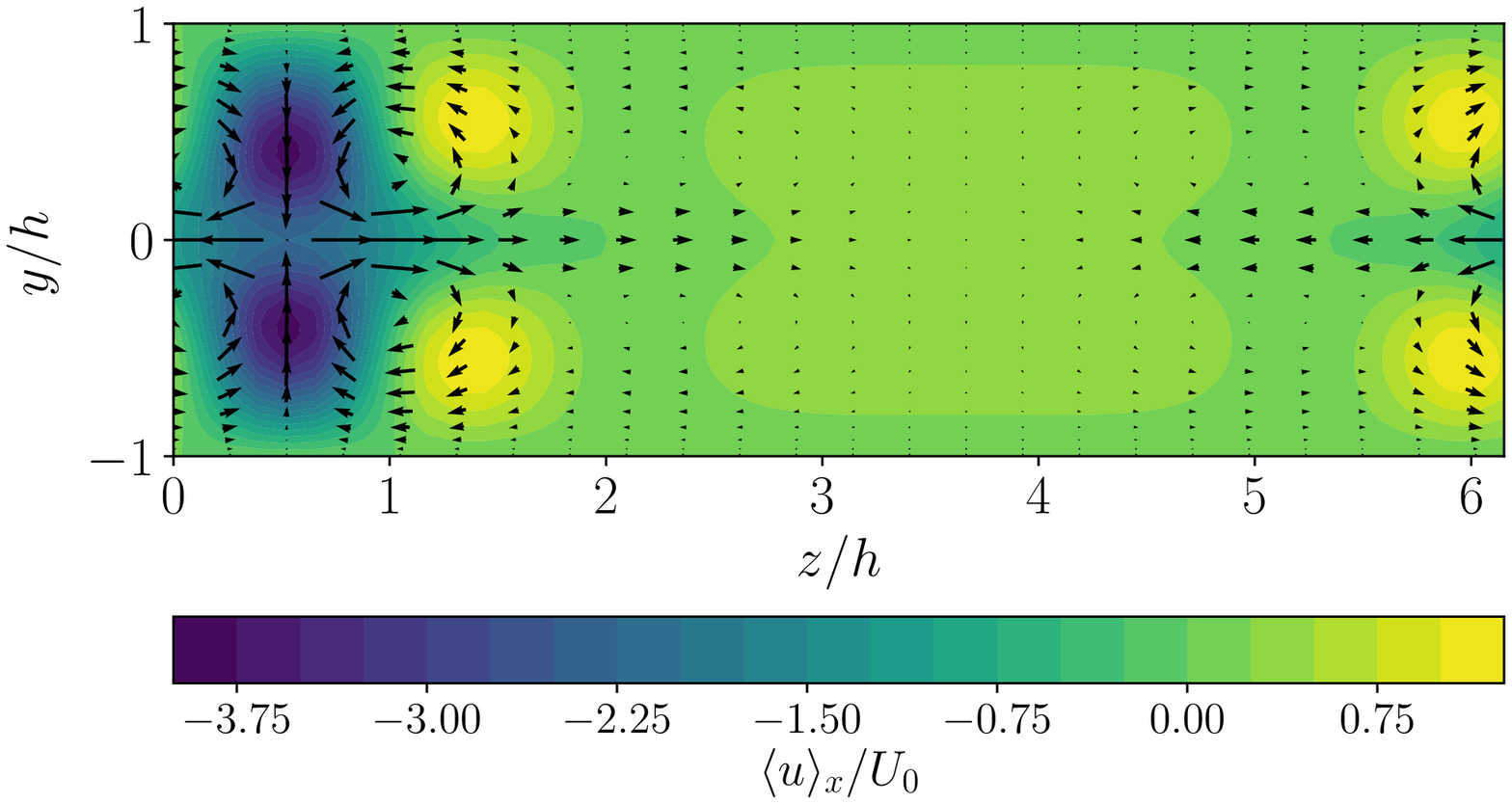}} 
	{\includegraphics[width=0.8\columnwidth]{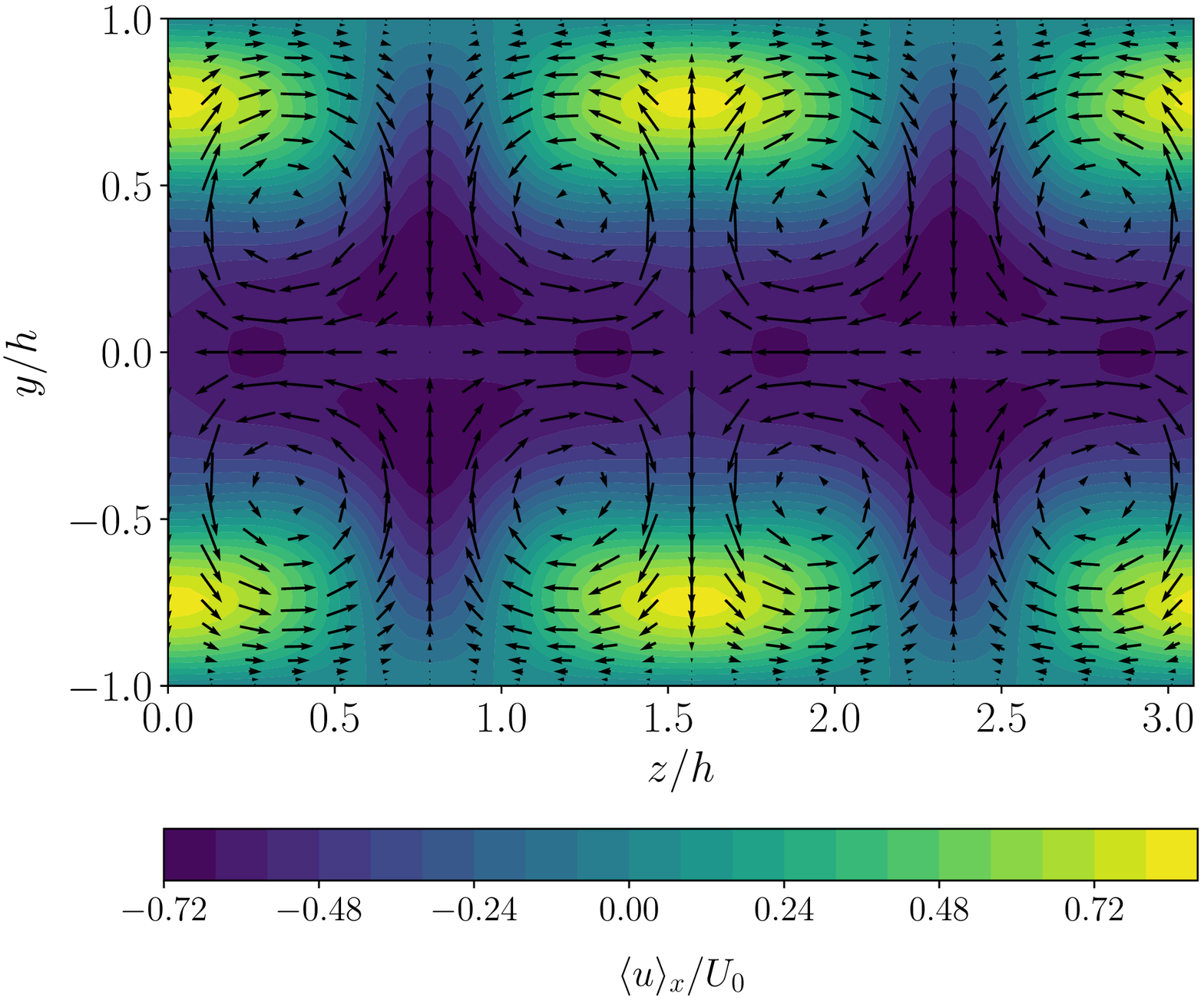}} 
\caption{
	Visualisation of the unstable eigenmodes of TW1 (top) and TW-sym (bottom) 
	showing the deviation of the streamwise average of 
	the streamwise velocity component, $\langle u \rangle_x$ from the laminar profile.
	The crossflow $(v,w)$ is indicated by the arrows.
}
\label{fig:visu-ef1}
\end{figure}

The structures differ in their spatial localisation and their degree of symmetry.  
The first one, TW1, has been calculated at $Re_0 =
1394$ in a domain of size $L_x/h \times L_y/h \times L_z/h = 2 \pi \times 2
\times 2\pi$ \citep{Zammert2014} and is an edge state in the full space.  
It is localised in the spanwise direction,
with two low-speed streaks accompanied by four vortices and is mirror-symmetric
about the midplane.  The second one, TW-sym, has been obtained by a
Newton-Krylov search at $Re_0 = 1010$ from an ECS in the domain $L_x/h \times
L_y/h \times L_z/h = 2 \pi \times 2 \times \pi$ that had originally been
calculated with constant flow rate \citep{Zammert2015}.  It consists of two
high-speed streaks, four low-speed streaks and eight vortices and is not
localised in the spanwise direction. 
Visualisations of the streamwise-averaged
structures and their respective leading unstable eigenmode 
are presented in figs.~\ref{fig:visu-edge} and \ref{fig:visu-ef1}, respectively, 
where the colour-coding represents the streamwise velocity component and the 
superimposed arrows the cross-flow. 

TW-sym is an edge state in a \blue{symmetry-invariant} subspace, 
\blue{that is, a subspace invariant under the action of a symmetry group,} only. 
Calculations of TW-sym are therefore carried out in a subspace that enforces 
mirror-symmetry about the midplane ($y=0$) and in spanwise direction 
about the plane $z = \pi/2$
\begin{align}
	\label{eq:symm-y}
	s_y\left((u,v,w)^t(x,y,z)\right) = (u,-v,w)^t(x,-y,z) \ , \\
	\label{eq:symm-z}
	s_z\left((u,v,w)^t(x,y,z)\right) = (u,v,-w)^t(x,y,-z) \ , 
\end{align}
where the superscript denotes the transpose.  Invariant solutions obtained in
\blue{symmetry-invariant} subspaces are also solutions with respect to the unrestricted
dynamics, where the number of unstable directions is usually higher
\citep{Duguet08,Kreilos2012,Avila2013}. In this context it is therefore of interest
to assess the effect of \blue{symmetry-invariant} calculations on feedback
stabilisation. For this reason, we also carried out controlled simulations of
TW1 within its \blue{symmetry-invariant} subspace. 
\blue{More precisely, the symmetry-invariant subspaces here are subspaces of the full domain  
invariant under the transformations defined in eqs.~\ref{eq:symm-y} and \ref{eq:symm-z} for TW-sym or, 
in case of TW1, in eq.~\ref{eq:symm-y} only.
}

\section{Stabilisation}
\label{sec:stabilisation}

\subsection{Pressure-based control}
\begin{figure}
	{\includegraphics[width=0.48\columnwidth]{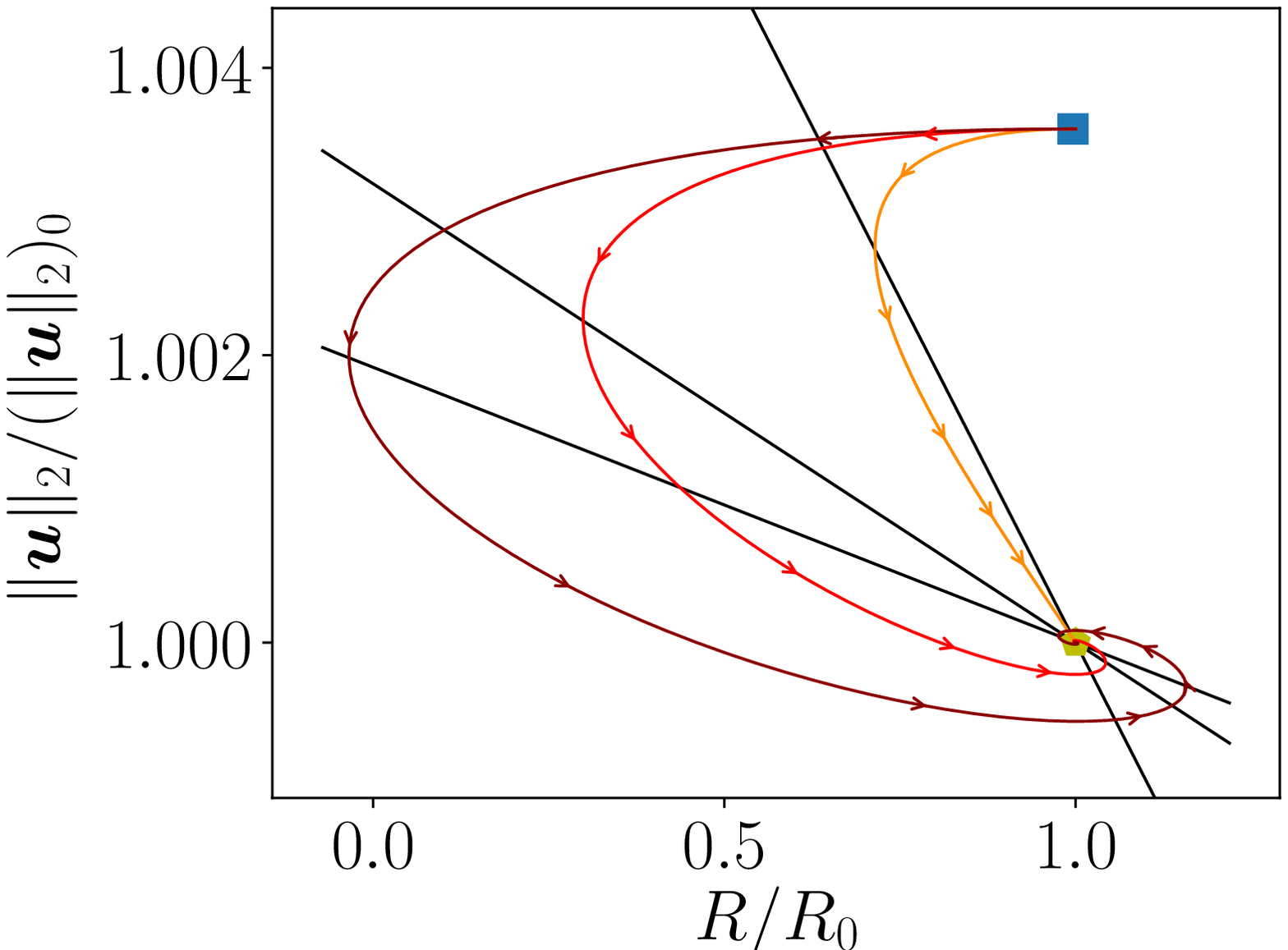}} 
	{\includegraphics[width=0.48\columnwidth]{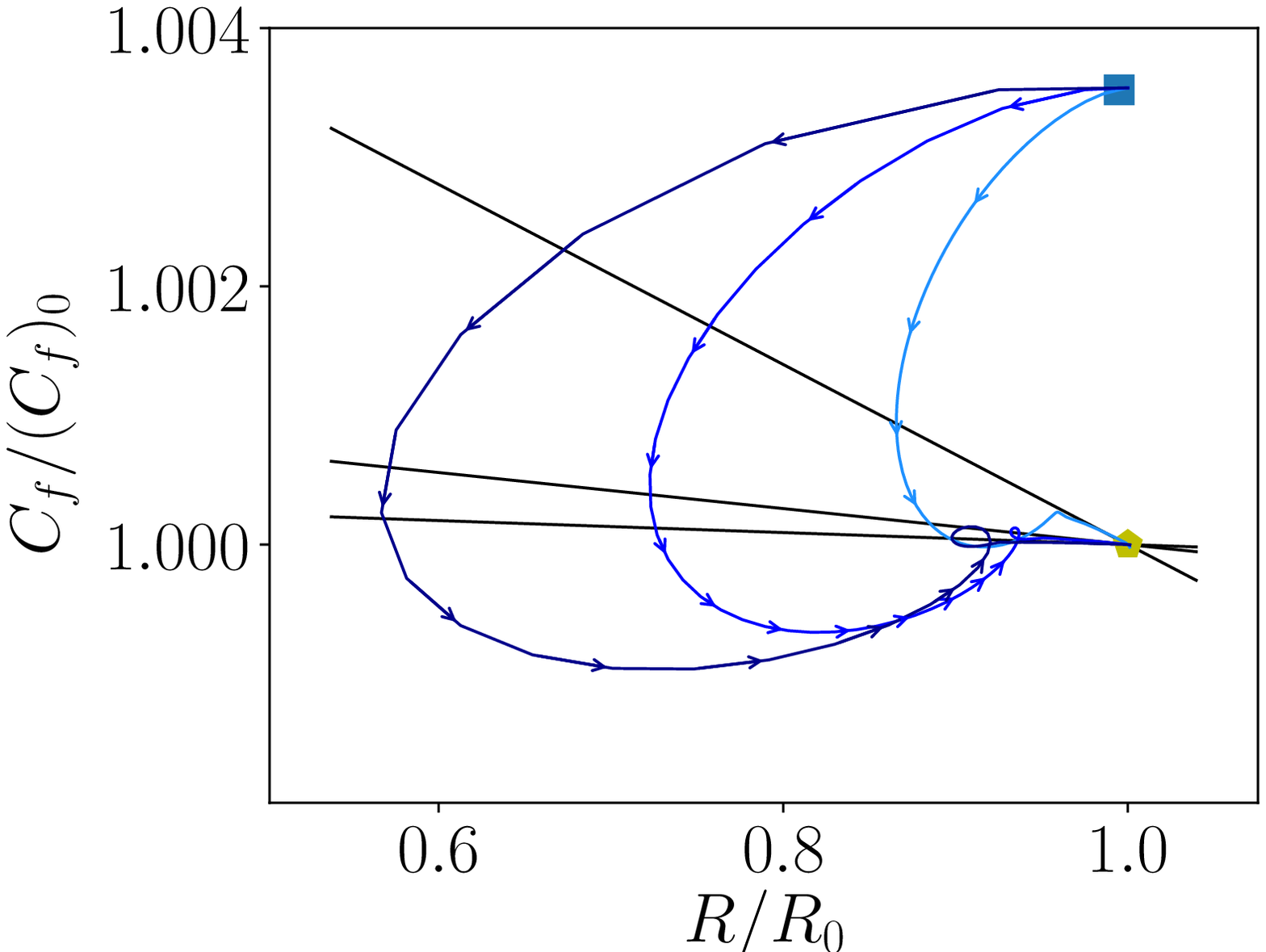}} 
	{\includegraphics[width=0.48\columnwidth]{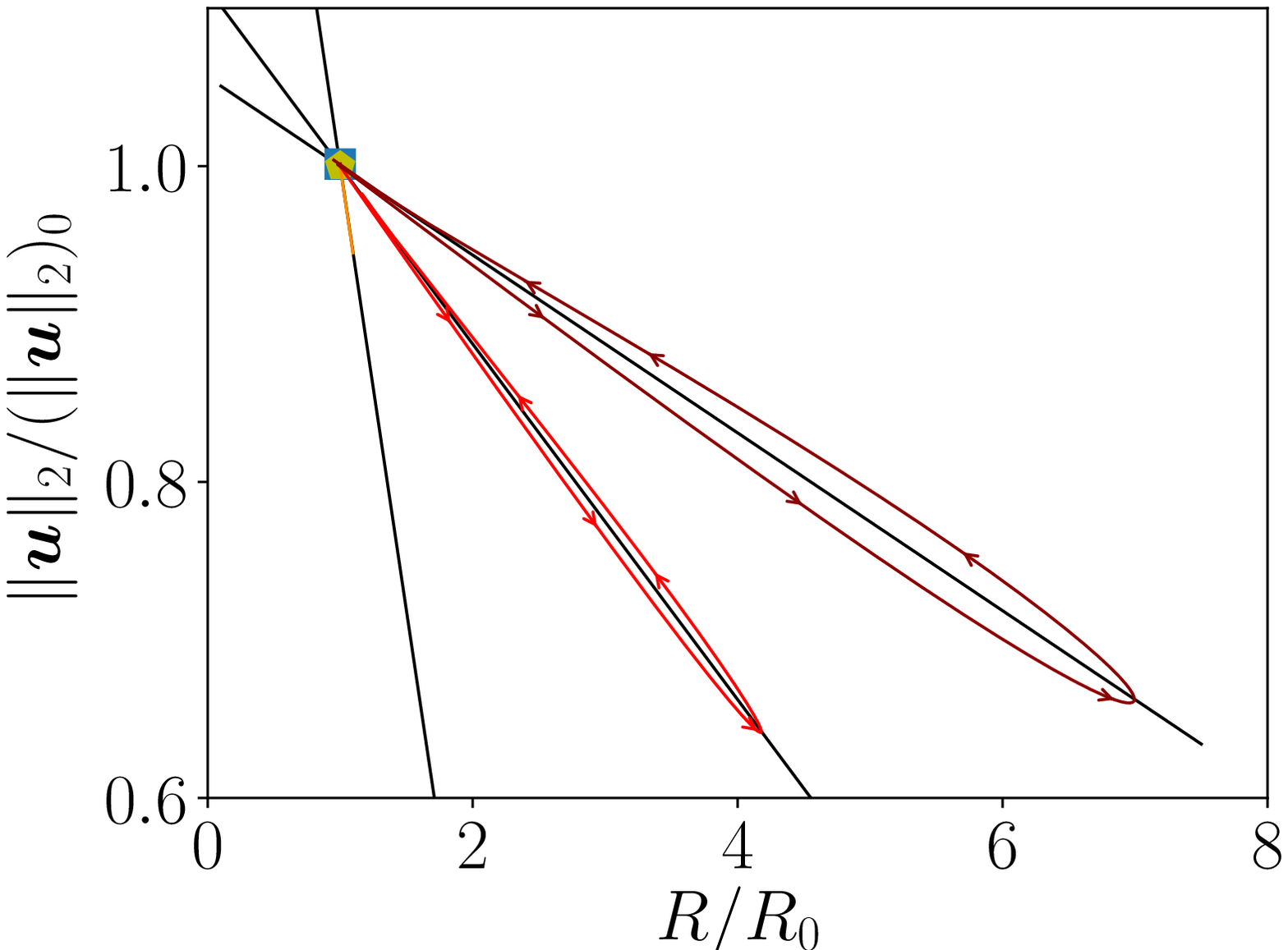}} 
	{\includegraphics[width=0.48\columnwidth]{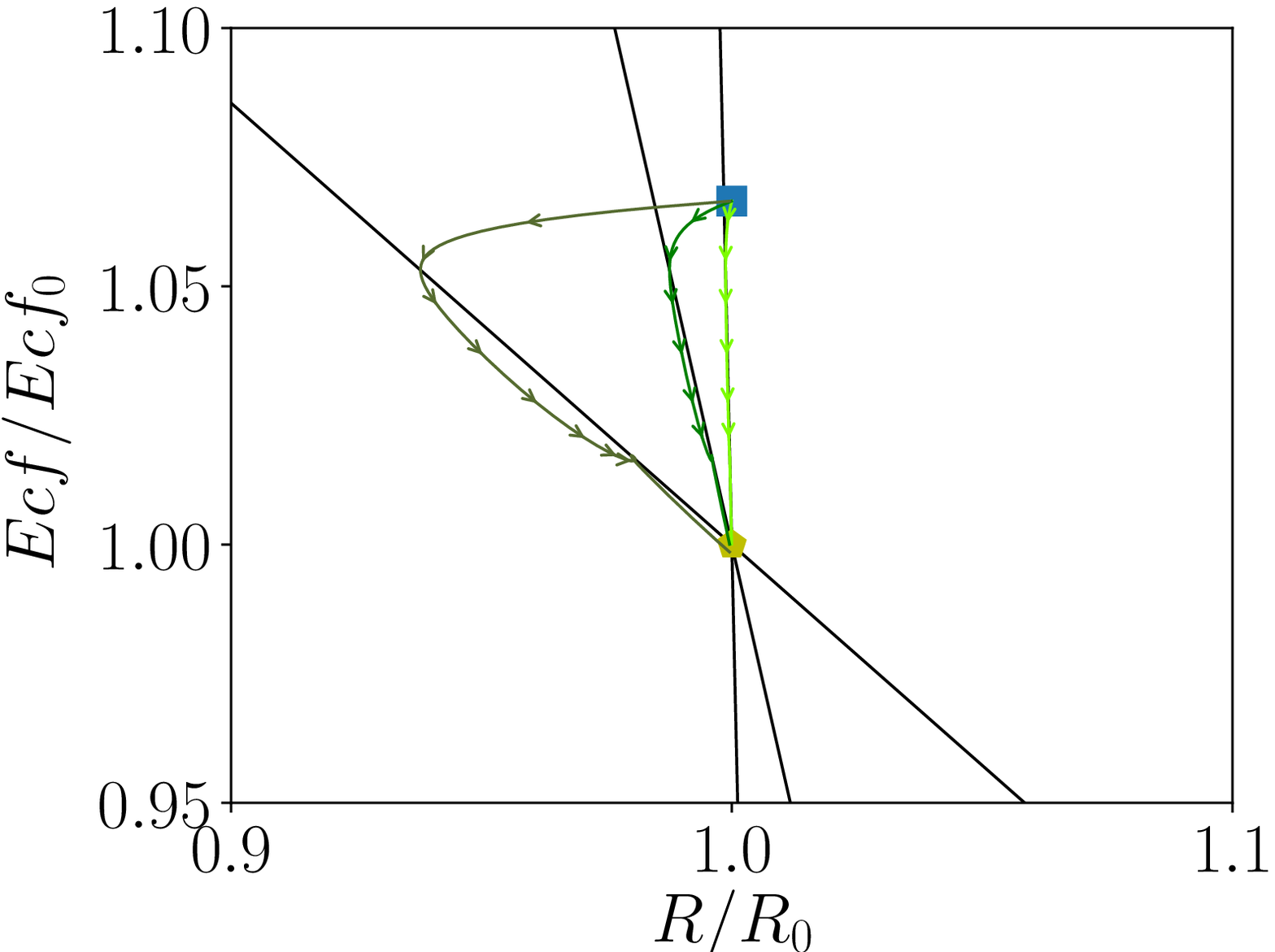}} 
\caption{
	Phase-space trajectories 
	for three different values of the control strength $\mu$ according to table \ref{tab:simulations} and 
different observables obtained from controlled DNSs according to 
eqs.~\eqref{eq:nse-dPdx1}-\eqref{eq:nse-dPdx4}.
Top left: TW1, $L_2$-norm; 
top right: TW1, friction factor $C_f$; 
bottom left: TW-sym, $L_2$-norm;
bottom right: TW-sym, cross-flow energy $Ecf$.
All calculations targetting TW-sym have been carried out in the 
	\blue{symmetry-invariant} subspace introduced in sec.~\ref{sec:numerics}.
}
\label{fig:phase-space-ppf}
\end{figure}

Figure \ref{fig:phase-space-ppf} presents phase-space trajectories of the controlled system 
for perturbations about TW1 and TW-sym with $\|\vec{u}\|_2$, the friction factor 
$C_f = 2\tau_w/(\rho U_0^2)$, where $\tau_w$ is the shear stress at the bottom wall, 
and the cross-flow energy 
\beq
Ecf(t) = \frac{1}{L_xL_yL_z}\int_0^{L_x} \int_{-L_y/2}^{L_y/2} \int_0^{L_z} dx dy dz \ \left(v^2(x,y,z,t) + w^2(x,y,z,t)\right) \ ,   
\eeq
as functions of the control parameter $R$, i.e. series TW1-A, TW1-B, TW-sym-A
and TW-sym-B in table \ref{tab:simulations}.  The top panels correspond to
results for series TW1-A (left) and TW1-B (right), and the bottom row for
TW-sym-A (left) and TW-sym-B.  All panels contain datasets from simulations
carried out with different values of the control strength $\mu$ indicated by
the colour gradient, where darker colours correspond to higher values of $\mu$
and hence stronger control. The corresponding control lines, which must
intersect at the operating point, are shown in black.  As can be seen from the
data shown in the two panels, the feedback control results in phase-space
trajectories where the perturbed edge state is driven towards the operating
point for all observables. In case of  TW1-A, the trajectories resemble those
from the model system discussed in sec.~\ref{sec:strategy} and shown in
fig.~\ref{fig:1D-model}. For the friction factor (TW1-B), the trajectories
first approach intermediate states on the control line and subsequently follow
the control line towards the operating point. For TW-sym-A the trajectories
show large excursions and eventually return to the operating point, while for
TW-sym-B the dynamics evolves along the control lines. 	 
We note that the trajectory passing through the point $R=0$ as in the top left panel of fig.~\ref{fig:phase-space-ppf} 
does not necessarily result in laminar flow. At this point the control input cancels the 
reference pressure gradient, resulting in an instantaneously vanishing production 
term for the deviations of the laminar profile. However, deviations from the laminar profile 
can still be present in the flow. Relaminarisation may occur if the time scale at which the 
control acts is much larger than the time scale for the free decay of the cross-flow.

\begin{figure}
\centering	´
	{\includegraphics[width=0.48\columnwidth]{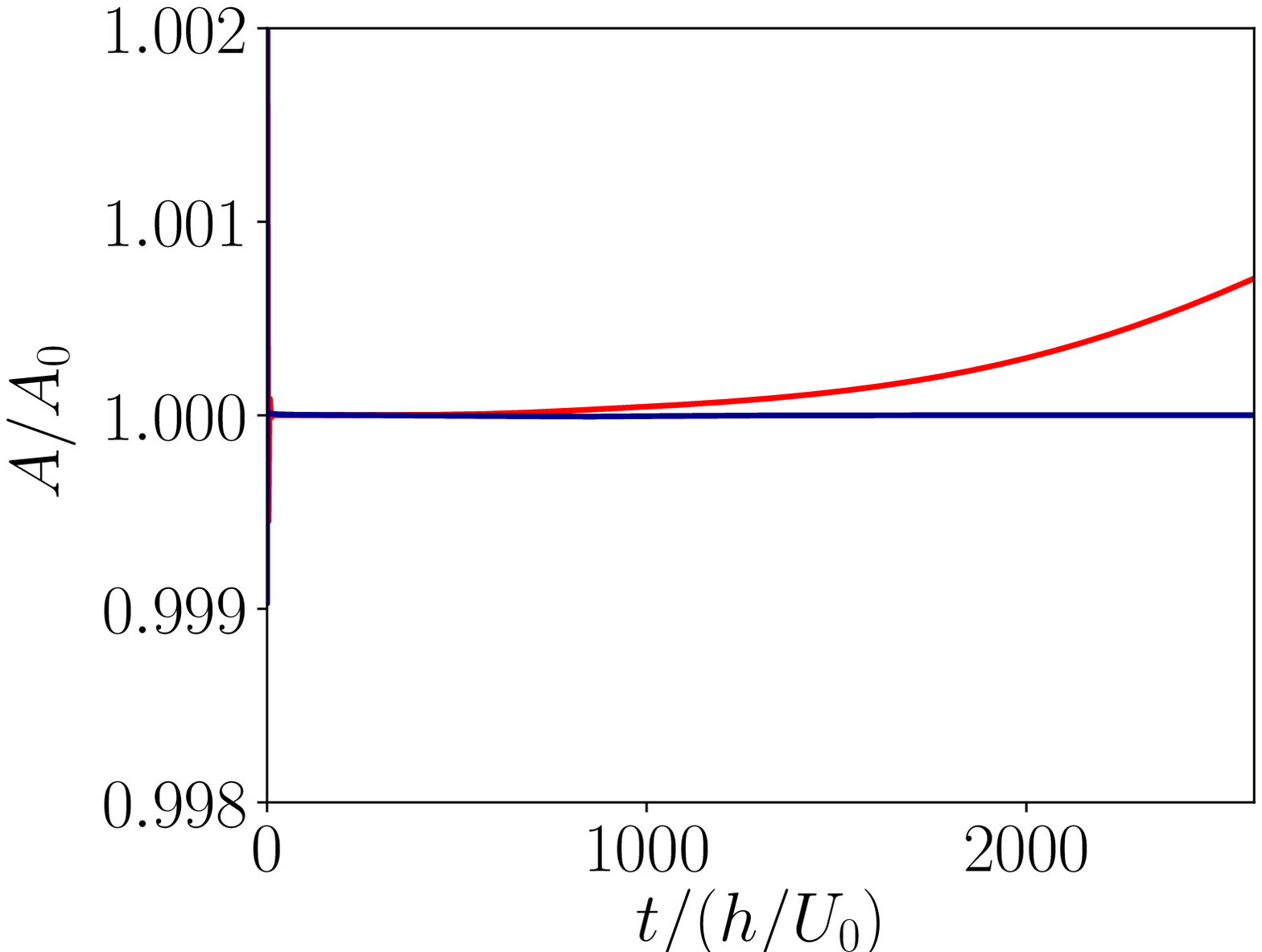}} 
	{\includegraphics[width=0.48\columnwidth]{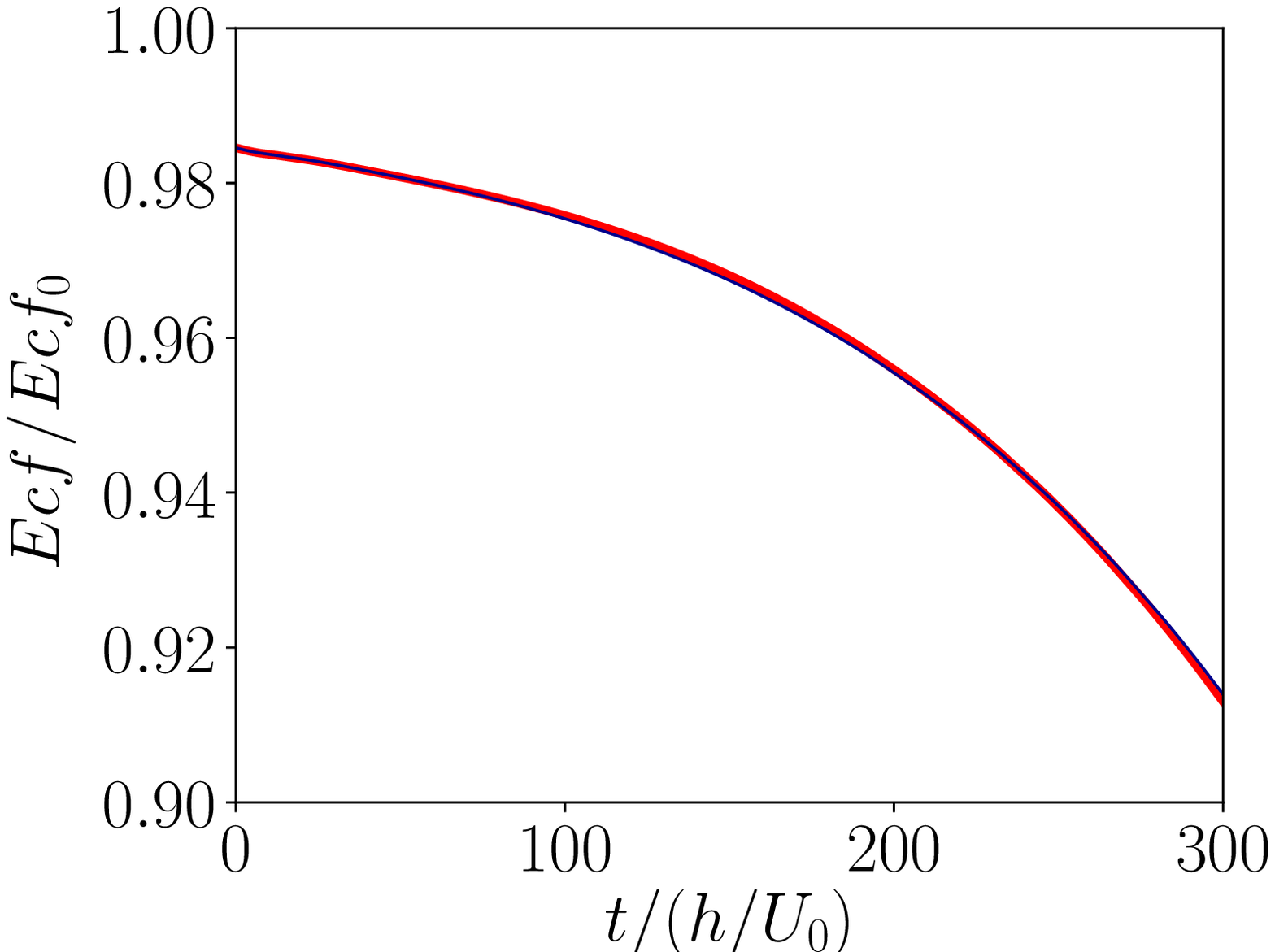}} 
\caption{
	Time evolution of the control observables (left) and the cross-flow energy (right) 
	for the pressure-controlled simulations with target state TW1. 
	Dynamics controlled with respect to the $L_2$-norm with $\mu = 10^6$ and with respect to 
	$C_f$ with $\mu = 3 \times 10^6$ are shown in red (grey) 
	and blue (dark grey), respectively. 
}
\label{fig:evolution-ppf}
\end{figure}

The simulations shown in fig.~\ref{fig:phase-space-ppf} reached close vicinity
of the operating point after very short simulation times 
(around $20$ time units for
norm-controlled simulations around $50$ time units for 
friction-controlled simulations) 
of both TW1 and TW-sym.  However, if the controlled system is evolved for very
long times, the trajectories leave the operating point again.  This
is demonstrated by the time evolution of $\|\vec{u}\|_2$ shown in red (light
grey) and $C_f$ shown in blue (dark grey) in the left panel of
fig.~\ref{fig:evolution-ppf} for the operating point TW1.  A deviation of
$\|\vec{u}\|_2$ from the reference value is visible after about 1000 time
units, while $C_f$ appears to remain constant.  The right panel of
fig.~\ref{fig:evolution-ppf} presents the time evolution of the cross-flow
energy
with $v$ and $w$ being the wall-normal and spanwise components of $\vec{u} =
(u,v,w)$.  
The control is
unable to prevent the dynamics from escaping from the operating point
towards the laminar fixed point. Interestingly, this happens on a much shorter
timescale compared to the departure of the control observables from their
target values.  Similar observations can be made for the dynamics of TW1
controlled with respect to $C_f$, for TW-sym and for controlled simulations of
TW1 carried out in its \blue{symmetry-invariant} subspace (not shown).  

The results shown in fig.~\ref{fig:evolution-ppf}
suggest the presence of a residual instability in the controlled simulations. 
According to the discussion in sec.~\ref{sec:stable-dir}, an instability in the
controlled system could result from the control being too weak to completely
remove the original instability, from the control being orthogonal to the
unstable direction as would be the case for strictly periodic instabilities, or
from an undesired destabilising effect of the control on the stable directions.
The first possibility can be ruled out by an exhaustive parameter scan. The
second possibility does not apply either, as the unstable directions have
non-zero streamwise mean as discussed in sec.~\ref{sec:operating_points} and
thus overlap with the control.  In what follows we therefore investigate in
detail how the control alters the tangent space structure of the chosen
invariant solutions.

\subsubsection{Effect of the control on stable and unstable directions}
\begin{figure}
	\centerline{\includegraphics[width=\columnwidth]{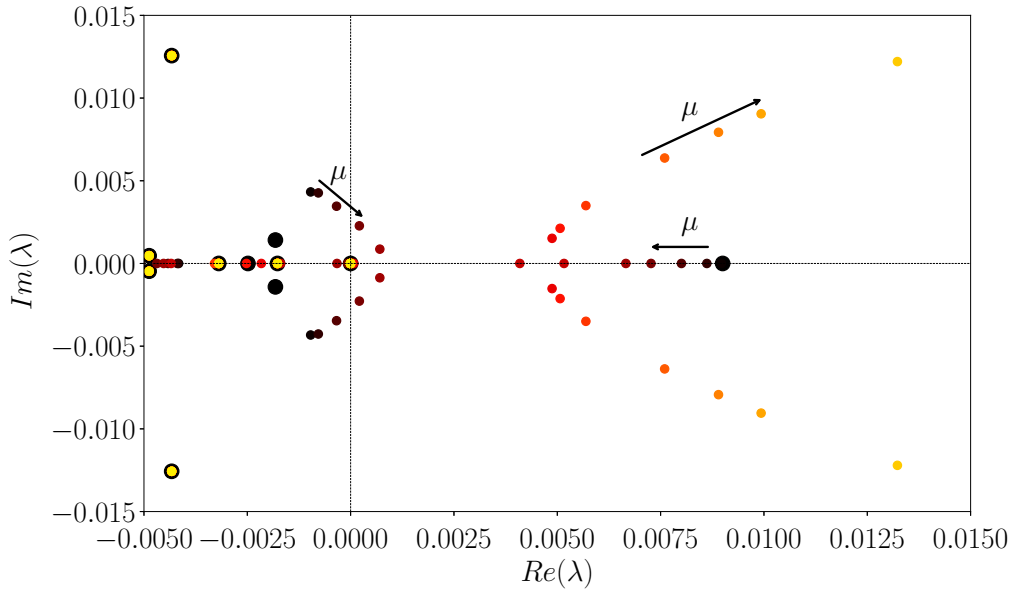}} 
\caption{
Spectrum of the Jacobian at TW1 for the combined system DNS with feedback control according to 
eqs.~\eqref{eq:nse-dPdx1}-\eqref{eq:nse-dPdx4} as a function
of the control strength $\mu$. The thick black dots correspond to the uncontrolled system 
and the decreasing color gradient indicates increasing values of $\mu$.
The positive real eigenvalue corresponding to the original instability 
decreases with increasing $\mu$ and eventually changes sign by jumping
from 0.005 to -0.0025 on the real axis.
With increasing $\mu$ a new feedback-induced instability occurs, represented by the 
complex eigenvalues with positive real parts.
}
\label{fig:ppf-eigenvalues}
\end{figure}

In order to quantify the effect of the control on the tangent space of the
invariant solutions investigated here, stability analyses of TW1 with respect
to the coupled system consisting of DNS and feedback control as in
eqs.~\eqref{eq:nse-dPdx1}-\eqref{eq:nse-dPdx4} have been carried out, see
series TW1-A-stab listed in table \ref{tab:simulations}.  Figure
\ref{fig:ppf-eigenvalues} shows the eigenvalues of the Jacobian at TW1 for the
free dynamics and for the $L_2$-norm controlled system for different values of
the control strength $\mu$.  As can be seen, the free dynamics is such that TW1
has one unstable direction as expected for an edge state. For low values of
$\mu$ the corresponding single positive real eigenvalue decreases with
increasing $\mu$. At the same time, a pair of complex conjugate eigenvalues
with negative real part move closer to the line where the latter vanishes.
Eventually, their real part becomes positive indicating the presence of a new
unstable direction.  For a small set of parameters, both old and new unstable
directions are present. That is, even though the original unstable direction is
removed for large enough $\mu$, the control indeed destabilises stable
directions of the uncontrolled system. 
The neutral directions associated with continuous shift symmetries remain unaffected, 
as can be seen by considering the eigenvalues with zero real parts in fig.~\ref{fig:ppf-eigenvalues}. 

\citet{Willis17} also calculated eigenvalues and Floquet exponents for their
successfully stabilised invariant solutions.  In both cases there are stable
eigenvalues whose real parts move closer to zero in the controlled system, see
fig.~3a of \citep{Willis17} for the spectrum of a travelling wave, and fig.~4c
for the Floquet exponents of a stabilised periodic orbit. In summary, 
a simple one-dimensional feedback control can have adverse effects on the stable directions, whereby the
real parts of the stable eigenvalues tend to zero and may even become positive
as shown here. This precludes the application of the pressure-based feedback control
to the search for new invariant solutions in channel flow following 
the procedure proposed by \citet{Willis17} for pipe flow, as without any information 
about eventual overlaps between the control and the stable directions it is 
difficult to know {\em a-priori} if such a feedback-induced instability indeed 
occurs. 
Hence a black-box application of such feedback strategies without good 
knowledge of the coefficients is not guaranteed to work. 
Before returning to this point in more detail in the following section, we
briefly discuss the experimental applicability of this method in terms of
turbulence control.  

\subsubsection{Potential experimental applicability}
Although the feedback control does not stabilise the
operating point,  it is able to find global target observables connected with
the streamwise component of the flow, which do not require knowledge of all
velocity components and are thus easier accessible experimentally.     This
suggests that the proposed pressure-based feedback control can be used to
confine turbulent dynamics to a region of phase space selected by a given value of e.g. the friction factor
and to prevent large fluctuations in kinetic energy or drag.  Preliminary
results for a wall-suction-based feedback control for plane Couette flow
\citep{Linkmann19c} show that this is indeed the case, at least in small domains. 
This suggests that further research into the effect of pressure-based linear feedback 
on the global properties of a flow may be worthwhile to pursue. In order to fully assess the 
potential experimental viability of such an approach, it is of paramount importance to 
carry out numerical simulations in domains with large streamwise extent, much beyond the minimal 
flow units used in the present study.

\subsection{Adjoint control}
\label{sec:revised-control}
\begin{figure}
	{\includegraphics[width=0.48\columnwidth]{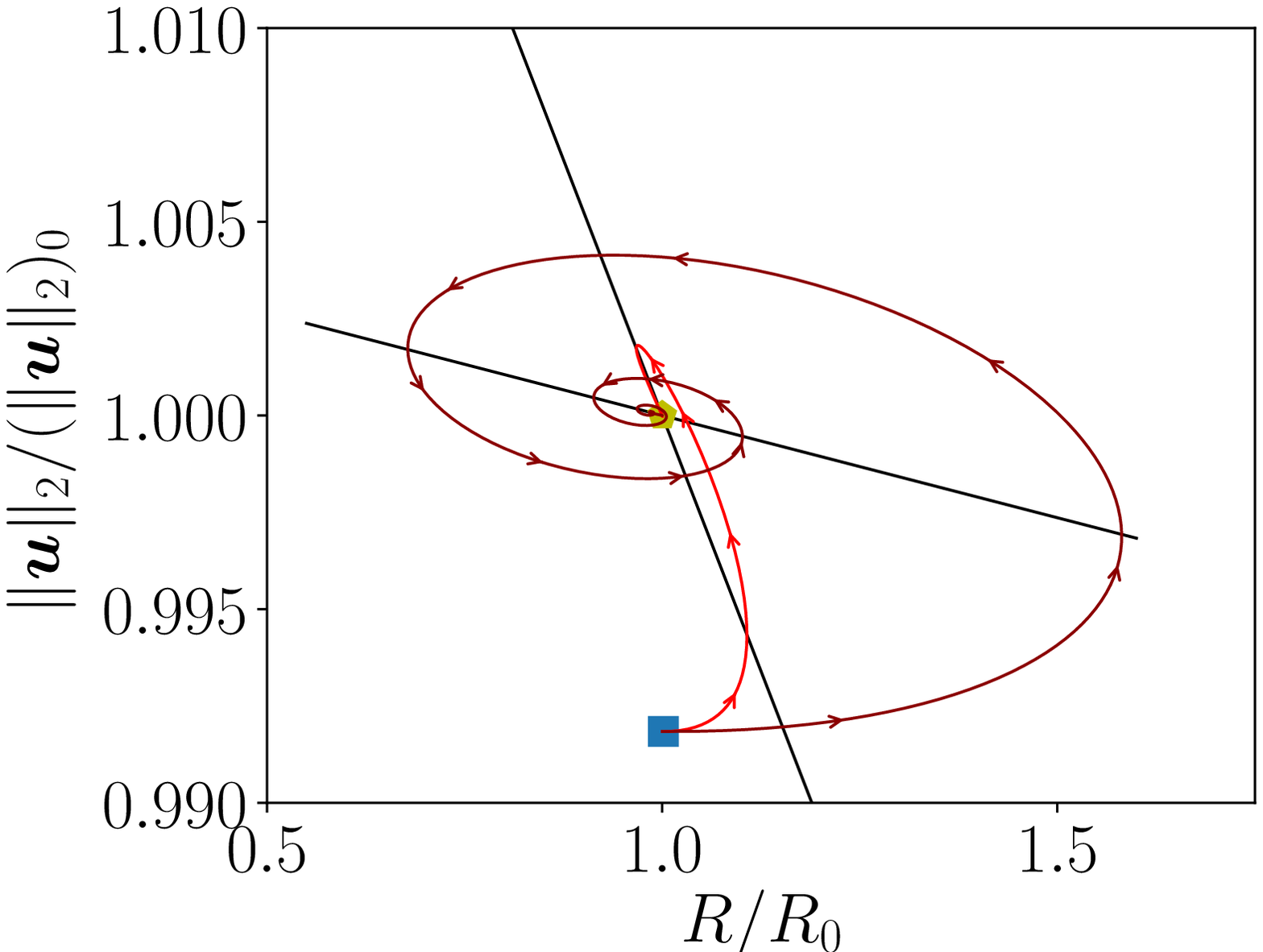}} 
	{\includegraphics[width=0.48\columnwidth]{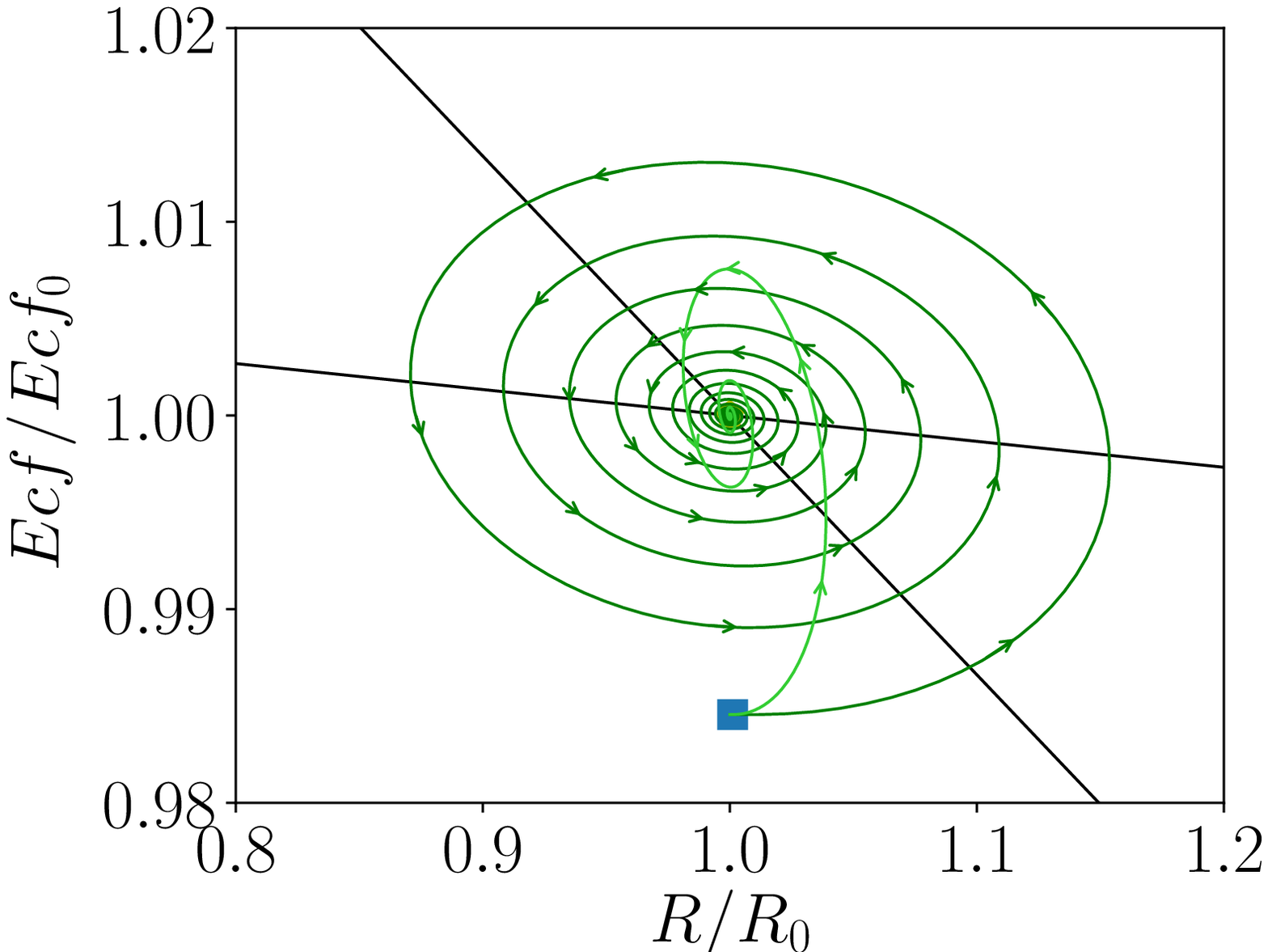}} 
	{\includegraphics[width=0.48\columnwidth]{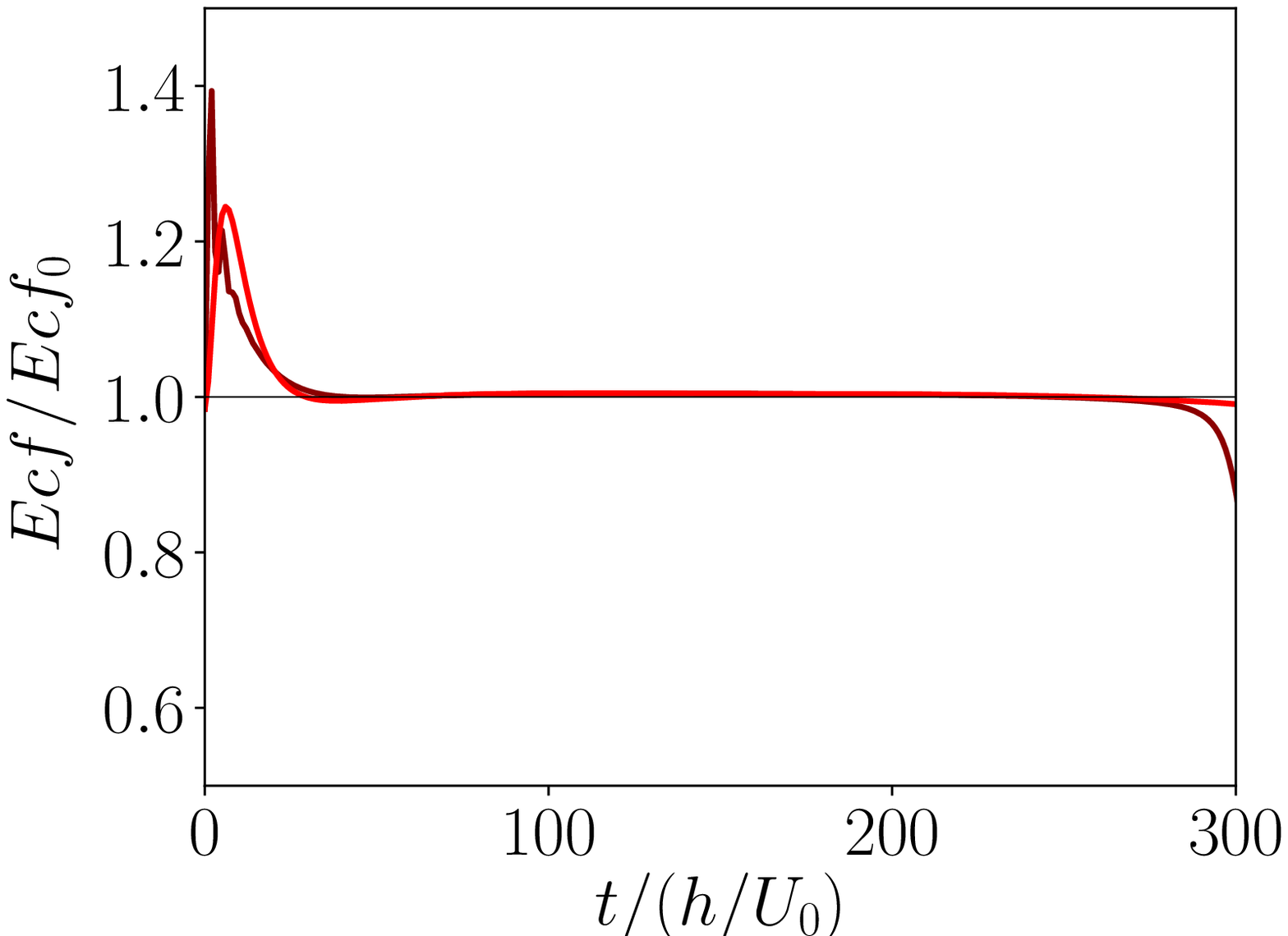}} 
	{\includegraphics[width=0.48\columnwidth]{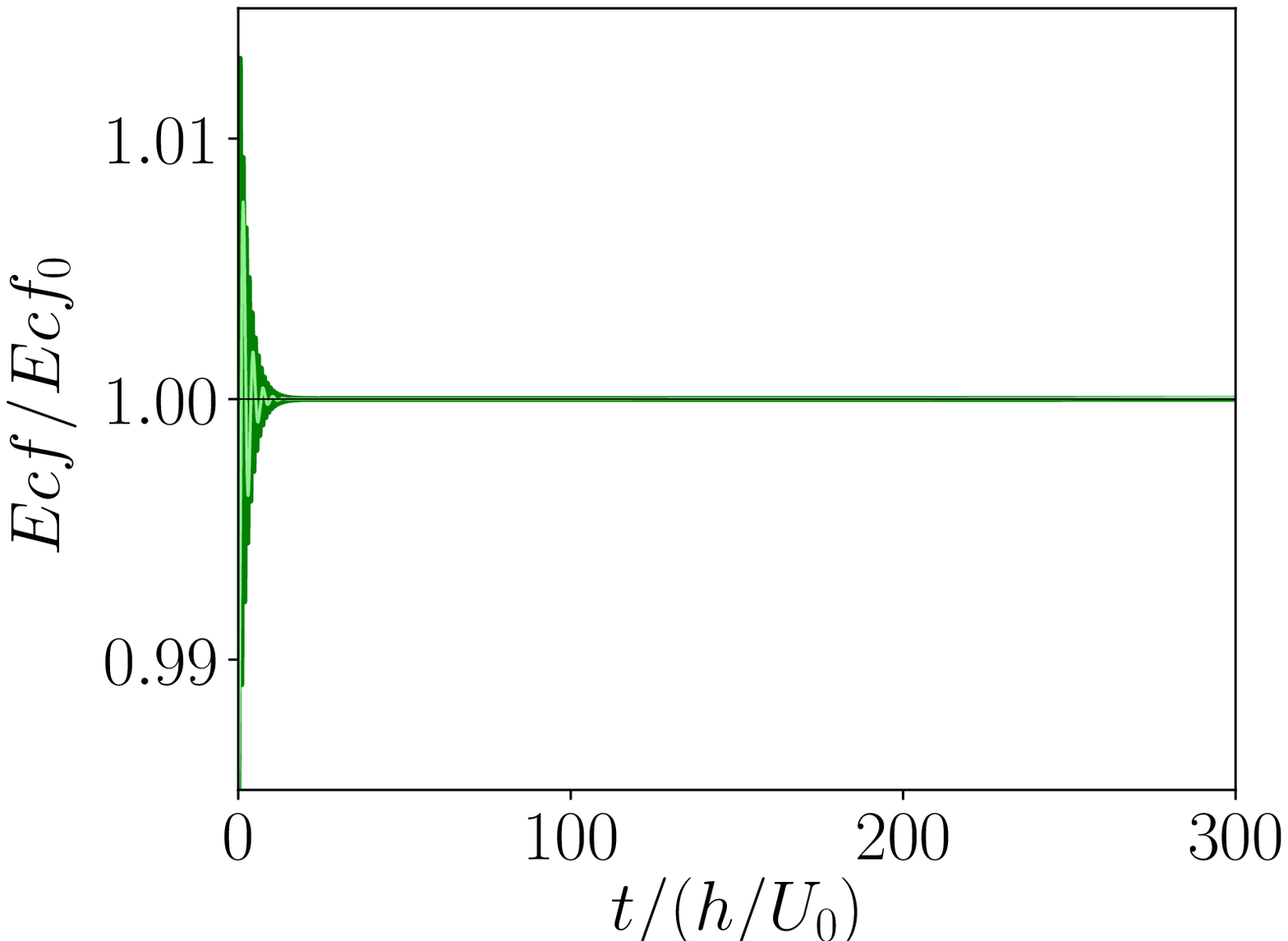}} 
\caption{
	TW1: Phase-space trajectories (top row) and corresponding evolution of the cross-flow energy (bottom row) 
	for two different values of the control strength $\mu$ 
	for the adjoint-based control procedure given by eq.~\eqref{eq:nse-dPdx1-dd} and eqs.~\eqref{eq:nse-dPdx2}-\eqref{eq:nse-dPdx4}
	with respect to the $L_2$-norm (left column, series TW1-C) and the cross-flow energy (right column, series TW1-D). 
	The control lines are indicated in black. 
}
\label{fig:phase-space-ppf-dualdir}
\end{figure}

Figure \ref{fig:phase-space-ppf-dualdir} shows phase-space trajectories with
respect to the $L_2$-norm and the cross-flow energy and time-series of the
latter obtained with the control implemented according to
eq.~\eqref{eq:nse-dPdx1-dd} and eqs.~\eqref{eq:nse-dPdx2}-\eqref{eq:nse-dPdx4}
for TW1, i.e.  series TW1-C and TW1-D summarised in table
\ref{tab:simulations}. The panels in the left column correspond to DNSs
controlled with respect to $L_2$-norm and the panels in the right column to
controlled runs with respect to the cross-flow energy.  As can be seen from the
phase space trajectories in the top-row panels, all controlled simulations
approach the targeted values of the chosen observables, as has been the case
for the streamwise-invariant control discussed in the beginning of
sec.~\ref{sec:stabilisation}.  The time evolution of the cross-flow energy
(bottom-row panels) indicates that the controlled system now also approaches
the actual operating point and stays in its vicinity for around 200 time units
for the $L_2$-norm control and for over 300 time units for cross-flow control.
However, eventually the state-space trajectory leaves the operating point again, as 
can be seen in the time evolution shown in bottom left panel of fig.~\ref{fig:phase-space-ppf-dualdir}, 
for instance.
The reason for this is most likely due to the choice of observable and thus with the control input. According to 
sec.~\ref{sec:adjoint_control}, the gradient of the input function $\kappa$ at the operating point
must be colinear with the unstable direction to achieve stabilisation. 
The results here suggest the presence of small overlaps. We will come back to this point later.

\begin{figure}
	{\includegraphics[width=0.48\columnwidth]{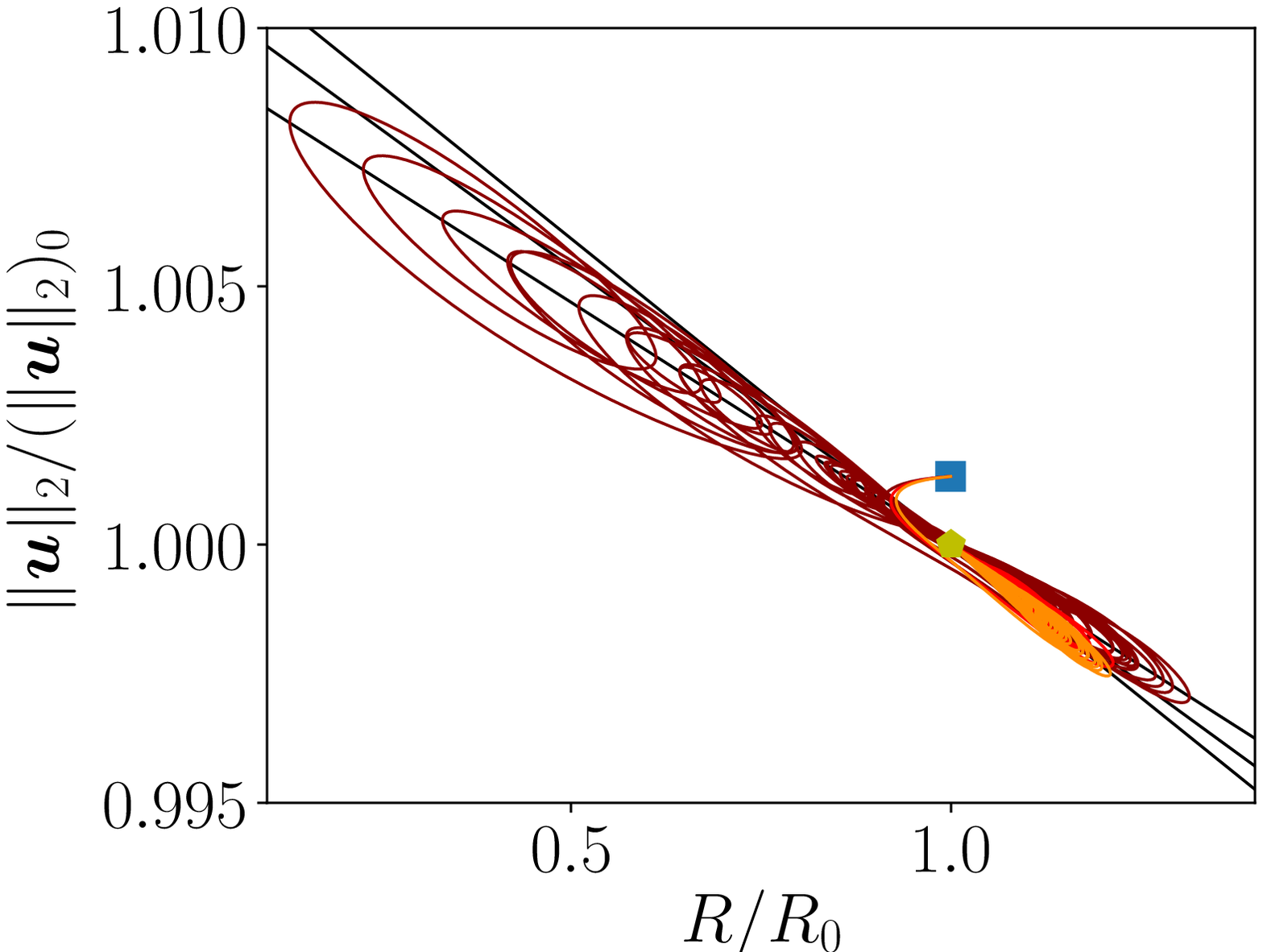}} 
	{\includegraphics[width=0.48\columnwidth]{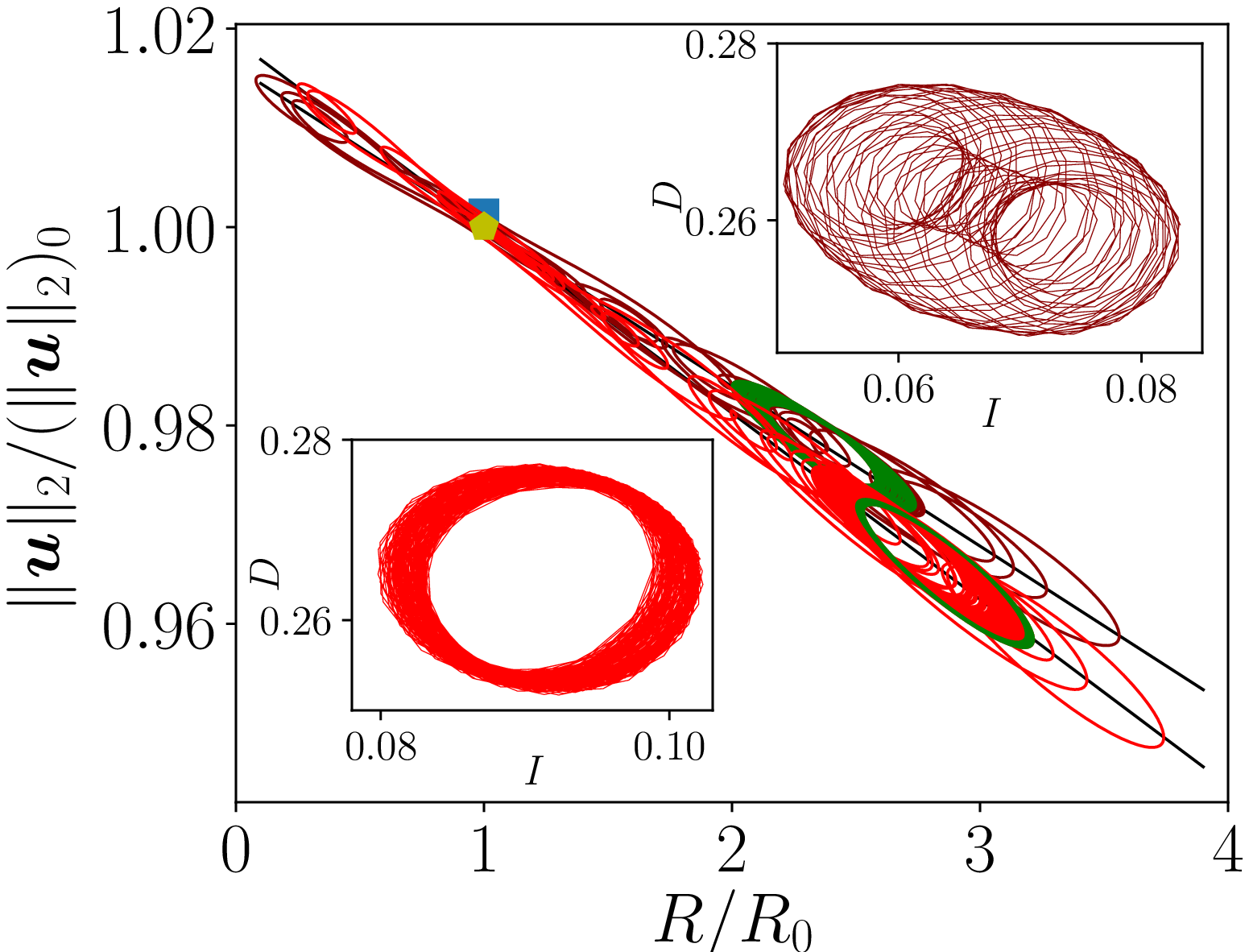}} 
	{\includegraphics[width=0.48\columnwidth]{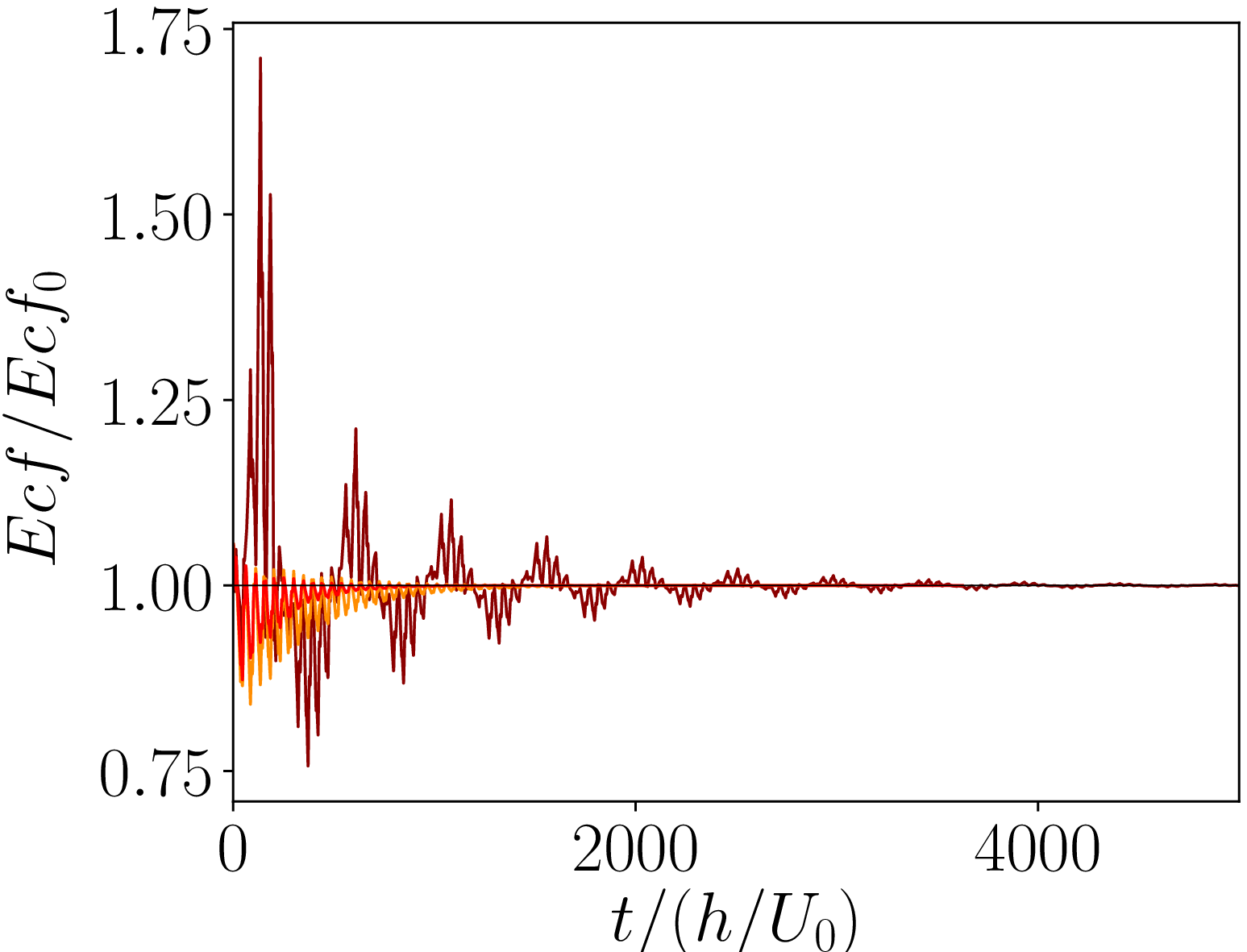}} 
	{\includegraphics[width=0.48\columnwidth]{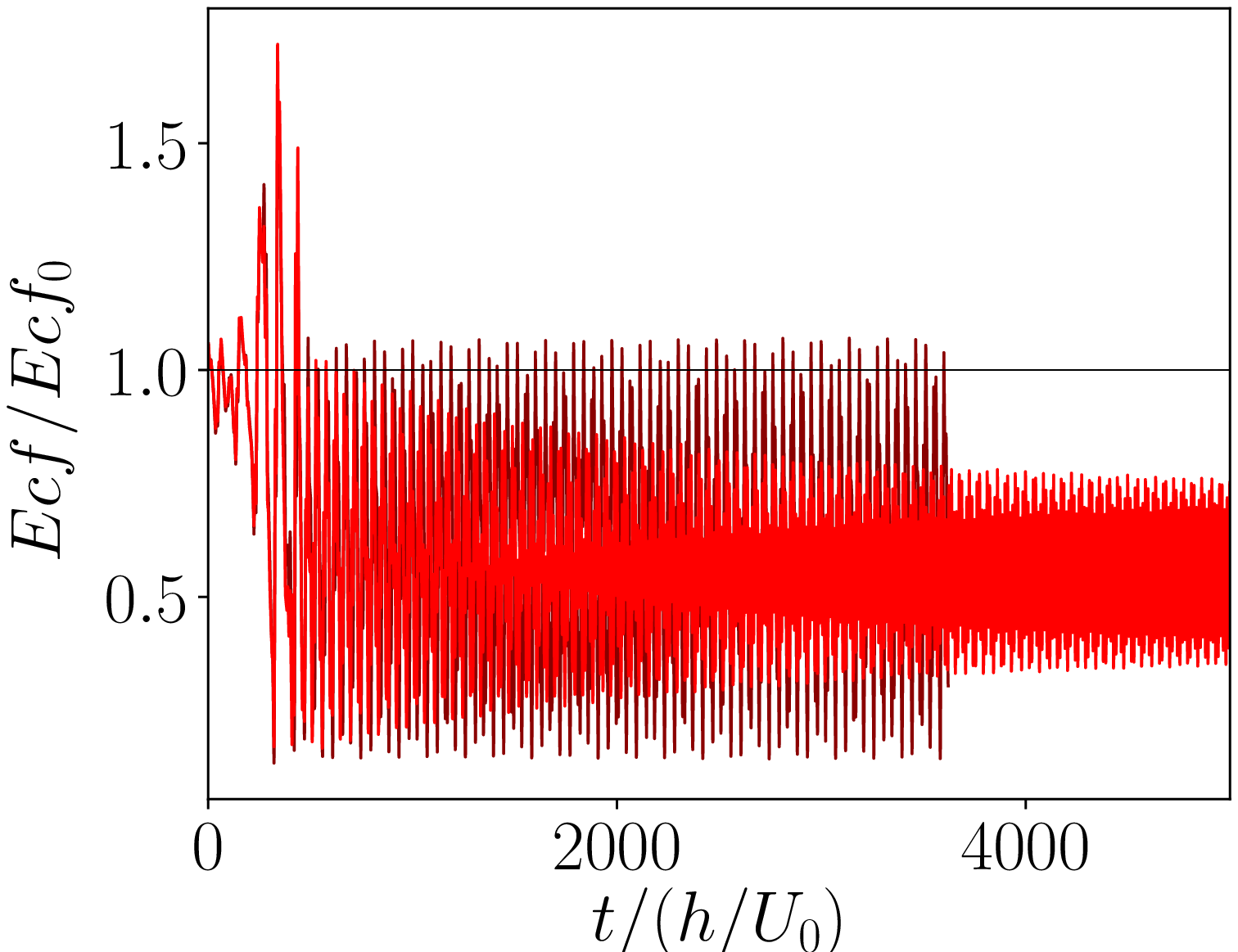}} 
	\caption{
	TW-sym: Phase-space trajectories (top row) and corresponding evolution of the cross-flow energy (bottom row) 
	for different values of the control strength $\mu$ 
	for the adjoint-based control procedure given by eq.~\eqref{eq:nse-dPdx1-dd} and eqs.~\eqref{eq:nse-dPdx2}-\eqref{eq:nse-dPdx4}
	with respect to the $L_2$-norm (left: TW-sym-C3-C5, right TW-sym-C1-C2 of table \ref{tab:simulations}).  
	The green ellipsoids in the top right
	panel show new limit cycles/invariant tori in the 
	controlled dynamics of TW-sym-C1 and TW-sym-C2. These are also shown in the insets in energy input ($I$) - dissipation ($D$) coordinates.
	The control lines are indicated in black. All calculations have been carried out in the 
        \blue{symmetry-invariant} subspace introduced in sec.~\ref{sec:numerics}. 
}
\label{fig:phase-space-ppf-dualdir-TWsym}
\end{figure}

Results from controlled simulations targetting TW-sym, 
all of which have been carried out in the \blue{symmetry-invariant} subspace introduced in sec.~\ref{sec:numerics}, 
that is, series TW-sym-C in table \ref{tab:simulations}, are
presented in fig.~\ref{fig:phase-space-ppf-dualdir-TWsym}.  Here, stabilisation
has been achieved using the $L_2$-norm as an observable, as can be seen from
the phase-space trajectories of runs TW-sym-C3, TW-sym-C4 and TW-sym-C5 in the
top left panel and the corresponding evolution of the cross-flow energy in the
bottom left panel of the figure.  Compared with the controlled dynamics
targetting TW1 carried out in the full space and shown in
fig.~\ref{fig:phase-space-ppf-dualdir}, the approach to the operating point is
much slower, but the stabilisation is complete.  For low values of the control
strength $\mu$, i.e. for runs TW-sym-C1 and TW-sym-C2, the controlled dynamics
gets trapped into new invariant tori, where the mean values of the $L_2$-norm
and the cross-flow energy depend on $\mu$ as shown by the phase-space plots and
the time evolution of the cross-flow energy in the right panels of
fig.~\ref{fig:phase-space-ppf-dualdir-TWsym}.  In all cases the phase-space
trajectories shown in fig.~\ref{fig:phase-space-ppf-dualdir-TWsym} remain in
the vicinity of the control lines, that is, the control procedure confines the
dynamics to regions phase space close to the chosen control lines.

According to the discussion in sec.~\ref{sec:adjoint_control}, the success of the adjoint method depends on 
the choice of feedback function $\kappa$, which in turn depends on the choice of observable.  
Concerning the choice of observable, several observations can be
made from a comparison of the visualisations of the ECS in
fig.~\ref{fig:visu-edge} and those of their respective unstable directions
shown in fig.~\ref{fig:visu-ef1}.  For both structures we note that the
cross-flow varies very little between the ECS and its unstable direction, while
clear differences are visible in the streamwise velocity component at least for
TW1.  This suggests that the cross-flow energy should work better than the
$L_2$-norm as a control observable for TW1, which is indeed the case, 
as can be seen by comparison of the bottom two panels of fig.~\ref{fig:phase-space-ppf-dualdir}. For
TW-sym the $L_2$-norm worked well. Finally, we note that stabilisation
through the adjoint-based feedback strategy could not be achieved using the friction
factor $C_f$ as an observable. Since $C_f$ is linear in the velocity field, its
gradient at the operating point is a constant vector and its dual hence not
orthogonal to all stable directions. 
The high degree of symmetry enforced by the calculations in the \blue{symmetry-invariant} 
subspace facilitates stabilisation, as it only allows instabilites that are shift-and-reflect
symmetric, as is the target state itself. As such, an overlap between the gradient of $\kappa$ at the operating point 
and the unstable direction is much easier to achieve, as they share the same symmetries. 
This sensitivity highlights some of the limitations of global one-dimensional feedback control 
to stabilise exact coherent structures.   

\begin{figure}
        {\includegraphics[width=0.48\columnwidth]{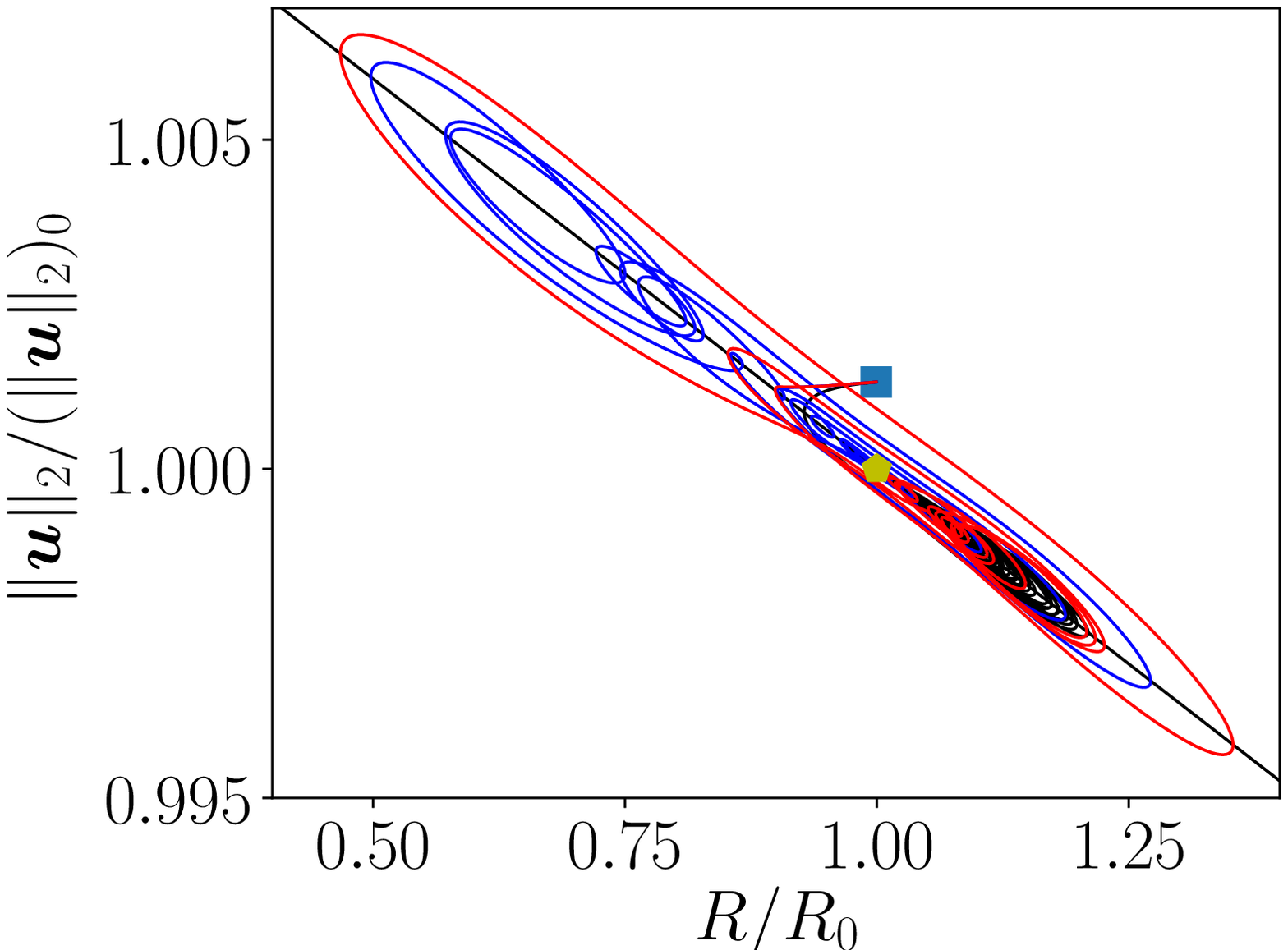}} 
        {\includegraphics[width=0.48\columnwidth]{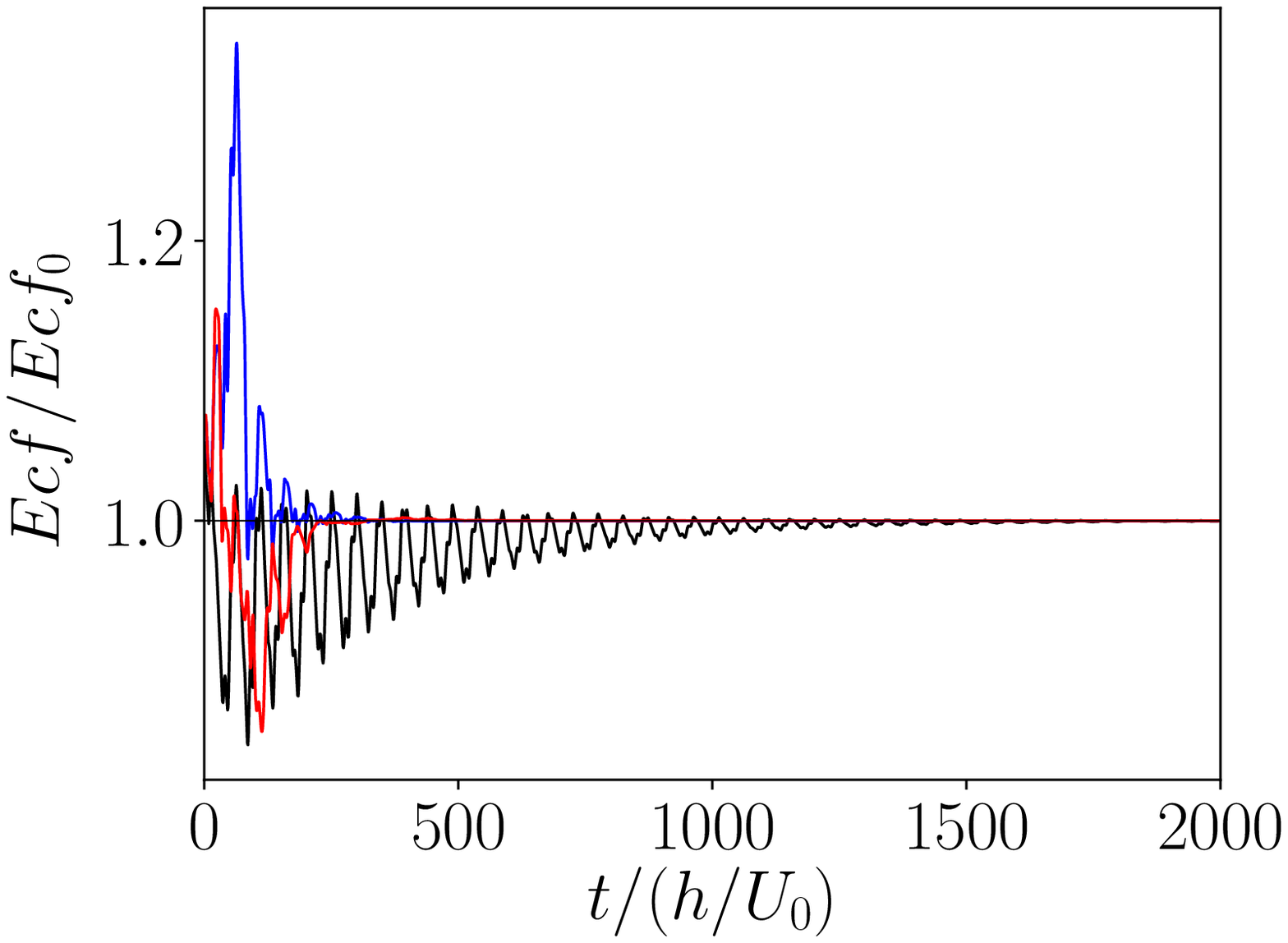}} 
        \caption{
	Phase-space trajectories (left) and corresponding evolution of the cross-flow energy (right)
        for TW-sym-C3 using the control procedure given by eq.~\eqref{eq:nse-dPdx1-dd} and eqs.~\eqref{eq:nse-dPdx2}-\eqref{eq:nse-dPdx4}
	with respect to the $L_2$-norm. The dual basis used in the control law has been calculated with respect to 40 (black), 
	50 (blue/dark grey) or 60 (red/light grey) stable eigenmodes. 
}
\label{fig:ppf-dualdir-TWsym-40-50-60}
\end{figure}

A few words on the performance limits, convergence and robustness of the control protocol are in order. 
Firstly, as the control procedure is based upon the linearised Navier\textendash Stokes equations, it is designed to work in
a neighbourhood of the operating point. In order to assess the performance limit of the
proposed controller, we carried out a parameter scan varying the magnitude of the random perturbation
$\delta \vec{u}$ at a fixed value of $\mu$. We found that the control protocol was successful for
$|| \delta \vec{u}||_2 / || \vec{u}^*||_2 < 0.25$, where $\vec{u}^*$ denotes the operating point (not shown).
Second, the dual unit vectors $v_e^*$ used in the calculations have been obtained
approximatively by calculating the dual basis of a subspace spanned by the
unstable eigenmode, the neutral eigenmodes and the first 40 stable eigenmodes, ordered by decreasing
Lyapunov exponent. For a smaller number of unstable eigenmodes, the controlled
dynamics did not recover edge state. Instead, it resembles that obtained for
weak values of the control strength shown in right panels of
fig.~\ref{fig:phase-space-ppf-dualdir-TWsym}, i.e. stable oscillatory
solutions of the dynamical system given by eq.~\eqref{eq:nse-dPdx1-dd} and 
eqs.~\eqref{eq:nse-dPdx2}-\eqref{eq:nse-dPdx4} were obtained.
Using a larger number of stable modes results in faster stabilisation as shown in 
fig.~\ref{fig:ppf-dualdir-TWsym-40-50-60} for example calculations of TW-sym-C3 using 
dual bases calculated with respect to 40, 50 and 60 stable eigenmodes. 
Third, calculations on finer grids require a more accurate calculation of the dual basis, i.e.                   
with respect to a higher-dimensional approximation of the solution's stable subspace.                
For instance, increasing the resolution from 48 to 64 Fourier modes in the homogeneous
directions required the dual basis to be calculated with respect to at least 80 stable 
directions to stabilise TW-sym (not shown).

\section{Discussion and conclusions}
\label{sec:conclusions}
In this paper we considered the application of linear feedback control as a strategy to 
stabilise invariant solutions of the Navier-Stokes equations. As an example of a canonical 
shear flow with a subcritical transition to turbulence, we considered channel flow at Reynolds numbers below the 
onset of linear instability at $\Rey = 5772.22$. We
focussed on the simplest possible problem, the stabilisation of edge states in 
minimal flow units, here travelling waves with one unstable direction.
Using explicitly constructed feedback strategies, the aim of the study was to point out and discuss 
the issues that arise when applying linear feedback control in attempts to 
stabilise exact coherent structures.
We devised two feedback control procedures. The first one is pressure-based, and thus in principle 
experimentally viable. The second one is constructed to remain orthogonal to the contracting and neutral
subspaces of the target state's tangent space. As it cannot be implemented in the laboratory, it mainly serves to 
highlight the complications that arise, in particular in comparison with the pressure-based method.
Simulations of the controlled systems were carried out for two target states that differ in their respective 
degrees of spanwise localisation. In case of the de-localised state, all calculations were 
carried out in its \blue{symmetry-invariant} subspace. 

The pressure-based control strategy was inspired by the work of
\citet{Willis17} on feedback stabilisation of edge states in pipe flow, where
the viscosity was adjusted as a function of energy-type observables. In order
to obtain a control procedure that, in principle, can be carried out
experimentally, we proposed to adjust the pressure gradient instead of the
viscosity as a function of either energy-type observables or the friction
factor.  
Even though the
control resulted in the dynamics approaching the respective target values of
the observables used in the control strategy, the actual edge states were not
stabilised as the control procedure has a destabilising effect on some of the
structures' contracting directions. 
This highlights that the success of similar methods such as that proposed by \citet{Willis17}
strongly depends on the system parameters and cannot be guaranteed to work in general. 
Stabilisation can be achieved if 
the control acts along the dual vector of the original unstable direction, that is, on a hyperplane
orthogonal to all stable and neutral directions.  
Here, it was found that care must be taken in the choice of observable, because the latter results in different
control input terms whose gradients may or may not overlap with the stable
directions. However, we found that for standard energy-type observables such
as the $L_2$-norm or the energy of the transverse fluctuations, only the de-localised and highly symmetric target 
state was stabilised, while for the spanwise localised target states the state-space trajectory of 
the controlled system remained very close to the target state for an extended time interval.
This emphasises the limitations of global one-dimensional feedback in the present context.

Apart from the observations summarised above, a few further issues deserve further attention in this context as they present 
obstacles that need to be addressed when designing linear closed-loop control strategies for the stabilisation of 
exact coherent structures. 

First, domains with periodic directions allow continuous 
symmetries in form of shifts along these directions, resulting in neutral modes with zero mean given   
by the derivatives of the target state in these directions, see e.~g.~\citep{Wolfe2006}. 
For the specific pressure-driven feedback this complication does not occur as the 
pressure gradient is constant 
in both streamwise and spanwise direction and therefore orthogonal to all modes with zero mean.
However, generally speaking, this issue needs to be taken into account in the design of 
linear control strategies applied to systems with continuous symmetries, in particular as 
neutral modes can be quickly destabilised by the control.
It may arise, for instance, in control strategies that directly modulate the flow. 
The adjoint-based method takes care of this problem by construction, with the important drawback that they can only be 
used in numerical simulations. 
This raises the general question as to how to design a 
practically relevant control strategy in systems with continuous symmetries that either leaves the neutral subspace 
unaffected or stabilises also the neutral modes of the uncontrolled system.

Second, a successful control strategy should in principle be applicable for a range of Reynolds numbers.
Considering specifically the stabilisation of edge states in the wider context of turbulence
control, the fact that edge states in plane Poiseuille flow disappear at Reynolds numbers 
above the laminar stability theshold limits the applicability
of the methods proposed here to unsteady, but not turbulent, channel flow at subcritical Reynolds numbers. However, 
similar complications arise also for shear flows like pipe or plane Couette flow where the laminar profile is linearly stable at all $\Rey$.
An important challenge for the application of linear closed-loop control to stabilise invariant solutions 
is connected with the contraction of the basins of attraction. 
In pipe or plane Couette flow, the basin of attraction of the laminar profile
contracts with increasing $\Rey$. A similar effect occurs for a stabilised exact coherent structure, its basin of attraction will 
contract with increasing $\Rey$ and domain size. For the controlled system, the situation is even more challenging as 
the increasing degree of instability requires higher feedback gain, resulting potentially in a further contraction 
of the basin of attraction of the invariant solution. 

Third, flow control is ultimately focussed on questions of practical relevance. In the present context, this
includes the combination of feedback control with classical methods for finding and continuing invariant solutions.
For linear control to be practically relevant, two conditions have to be satisfied: 
(i) the target state has to lie in the ergodic region of state space,
i.e., it should be approached closely by turbulent trajectories of the open-loop flow and (ii)
turbulent trajectories of open-loop flow should approach the target state to within a distance
smaller than the size of its region of attraction for closed-loop flow on practically accessible time
scales.  

Having discussed the challenges and limitations of global one-dimensional feedback in the context of invariant solutions, 
we now briefly mention applications where such strategies would (i) either be applicable as they are or with minor modifications, or 
(ii) where stabilisation of specific invariant solutions would be very useful.

Even though the pressure-based method fails as a means to stabilise invariant solutions, 
it is shown to target set values of the $L_2$-norm, the cross-flow energy or the friction factor. As such, 
pressure-based dynamic feedback may be a useful tool to accelerate or prevent
relaminarisation events.
Since feedback strategies alter the stability of an exact solution to the
Navier\textendash Stokes equations, they can not only be used to stabilise an operating
point, but also to further {\em destabilise} it, if so desired, or to confine
the dynamics to a certain region in phase space.  This may be useful in systems
where there is an interest in avoiding certain flow states, e.g.~those with
enhanced drag. Here, the issues discussed earlier are mitigated by the 
fact that the controller only needs to be efficient when the state-space 
trajectory approaches a small neighbourhood of the undesired state. 

Recent results from numerical simulations of channel flow suggest that extreme
fluctuations in the streamwise component of the wall-shear stress are less
likely if the flow is maintained by presciption of a constant flow rate
compared with forcing through a constant pressure drop or a fixed energy input
\citep{Quadrio2016}, however, the differences concerned rare events.  Dynamic
feedback could be a possibility to avoid extreme fluctuations more effectively.
In particular for shear flows with a pair of exact coherent structures 
born in a saddle-node bifurcation, typical extreme events \blue{should} correspond to the 
state-space trajectory following the heteroclinic connection from the lower to the 
upper branch. Stabilising states on the lower branch is an efficient way to suppress 
such extreme events.
\citet{Farazmand2019} showed that extreme events in 2D Kolmogorov flow can be
avoided by dynamically regulating the dynamics of certain Fourier modes at the
driving scale. Preliminary results show that a variant of the pressure-based 
feedback strategy proposed here can be applied to damp transverse fluctuations in plane Couette
flow through adjustable wall suction \citep{Linkmann19c}. This calls for
further investigations using in particular the pressure-based control strategy. 

\section*{Acknowledgements}
With great sadness we report that Bruno Eckhardt passed away shortly before 
finalisation of this manuscript. We thank him for his astute scientific insight 
and thoughtful guidance. ML thanks Yohann Duguet for very helpful discussions 
and the anonymous referees for their constructive feedback that has significantly improved the 
quality of the manuscript. 

\section*{Declaration of Interests} 
The authors report no conflict of interest.

\bibliographystyle{jfm}
\bibliography{refs}

\end{document}